 \documentclass[doublecolumn,journal]{IEEEtran}
\usepackage[utf8]{inputenc}
\usepackage{cite}
\usepackage{graphicx}
\usepackage[cmex10]{amsmath}
\usepackage{amsthm,amsfonts,amssymb,bm}  
\usepackage{array}
\usepackage[normalem]{ulem}
\usepackage{mathtools}
\usepackage[shortlabels]{enumitem}
\usepackage{subfig}
\usepackage{comment}
\usepackage{soul}
\usepackage{color,graphicx}
\usepackage{multirow}
\usepackage{float}
\usepackage{makecell}
 
\usepackage{hyperref}

\usepackage{multirow}
\usepackage{multicol}
\newcolumntype{L}[1]{>{\raggedright\let\newline\\\arraybackslash\hspace{0pt}}m{#1}}
\newcolumntype{C}[1]{>{\centering\let\newline\\\arraybackslash\hspace{0pt}}m{#1}}
\newcolumntype{R}[1]{>{\raggedleft\let\newline\\\arraybackslash\hspace{0pt}}m{#1}}
 
\newcommand{\floor}[1]{\left\lfloor #1 \right\rfloor}

\newcommand{{\snr}}{\mathop{\rm snr}}
\newcommand{{\inr}}{\mathop{\rm INR}}
\newcommand{{\SNR}}{\mathop{\rm SNR}}
\newcommand{{\INF}}{\mathcal{I}}

\newcommand{\ex}[1]{\mathbb{E}\left[ {#1}\right]}

\renewcommand\appendix{\par
	\setcounter{section}{0}
	\setcounter{subsection}{0}
	\renewcommand\thesection{Appendix \Alph{section}}
}

\usepackage{colortbl}
\usepackage[table,dvipsnames]{xcolor}

\title{Flexible User Mapping for Radio Resource Assignment in Advanced  Satellite Payloads}
\author{
\IEEEauthorblockN{Tom\'as Ram\'{\i}rez\IEEEauthorrefmark{1}, Carlos  Mosquera\IEEEauthorrefmark{1}, Nader Alagha\IEEEauthorrefmark{2}} \\ \vspace{1em}
\IEEEauthorblockA{\IEEEauthorrefmark{1} atlanTTic Research Center, Universidade de Vigo, Galicia, Spain} \\
\IEEEauthorblockA{\IEEEauthorrefmark{2} European Space Agency Technical Research Center (ESTEC), Noordwijk, The Netherlands} \\
\IEEEauthorblockA{Email: \{tramirez,mosquera\}@gts.uvigo.es,
nader.alagha@esa.int}
}

\begin{document}

\maketitle

\begin{abstract}
This work explores the flexible assignment of users to beams in order to match the non-uniform traffic demand in satellite systems, breaking the conventional cell boundaries and serving users not necessarily by the corresponding beams yielding more power. The smart beam-user mapping is jointly explored with adjustable bandwidth allocation per beam,  and tested against different techniques for payloads with flexible radio resource allocation. Thus, both beam capacity and load are adjusted to cope with the traffic demand. The specific locations of the user terminals play a major role, although their complexity does not increase as part of the proposed scheme. 
Numerical results are obtained for various non-uniform traffic distributions to evaluate the performance of the solutions. The traffic profile across beams is shaped by the Dirichlet distribution, which can be conveniently parameterized, and makes simulations  easily reproducible. Even with ideal conditions for the power allocation, both flexible beam-user mapping and adjustable power allocation similarly enhance the flexible assignment of the bandwidth on average. Results show that a smart pairing of  users and beams provides considerable advantages in highly asymmetric demand scenarios,  with improvements up to 10\% and 30\% in terms of the offered and the minimum user rates, respectively,  in hot-spot like cases.

\end{abstract}


\section{Introduction}\label{sec:Intro}

  Traditional architectures of High Throughput Satellites (HTS) are based on a rigid allocation of communication resources, not suited for a non-uniform geographical traffic demand, giving rise to excessive or insufficient offered throughput in different terrestrial cells. In recent years, more elaborate Radio Resource Management (RRM) has become instrumental for emerging flexible payload architectures in HTS, by exploiting the available degrees of freedom at the payload to address the requested traffic demand. Thus, RRM for advanced and flexible payload architectures has given rise to several results in the literature, in general highly dependent on the specific technological platform. The goal of such strategies is  to cope with the traffic demand through capacity transfer among beams \cite{Ang2021}. For instance, the flexible allocation of bandwidth to beams is discussed in 
  \cite{Park12}, whereas adjustable power allocation is  investigated in \cite{choi2005,Qi15,Efen20}. The joint allocation of power and bandwidth was early analyzed as a 
   mechanism to meet a potentially non-uniform demand across beams in \cite{Lei10,Wang2014}, and has been addressed in more recent works  \cite{cocco2018,paris2019,lagunas2020flexible_jour}. In addition, the time domain is also explored through beam hopping solutions in \cite{anzalchi2010,Lei11,Alberti10,lei2020beam,Xu2020}, among others. In all these works, it is a common assumption that users  are only served by their dominant beam, resulting in a rigid beam-user mapping. By blurring the conventional cell boundaries, additional flexibility emerges for the RRM process to offload traffic from more heavily demanded beams. Along this line of thought, load transfer is applied in  \cite{Ang2021} through  adaptive beamforming, by shaping the footprint of the different beams to the traffic profile, thus mitigating the potential demand-supply mismatch. Under the proposed change of paradigm in this paper, the association between users and beams is not fixed, which in conjunction with a flexible payload results in a hybrid approach, with both capacity and load transfer across beams. This  boundaryless concept for satellite systems has been previously studied for precoding solutions in \cite{Vaz18,Vaz20}, and for Power Domain NOMA (PD-NOMA) applications in  \cite{ICASSP20}. However, the flexible beam-user mapping remains unexplored in the context of RRM. 

 Even with traditional satellite payloads, a non-canonical beam-user mapping can prove to be particularly useful when 
serving hot-spots \cite{Nader2017,2018ICSSC,Tar18}. A hot-spot is, in general, a limited geographical area where the traffic is significantly larger than in the surrounding area,  putting significant strains on the satisfaction of the traffic demand 
 \cite{Icolari2017}. In the works \cite{Nader2017,2018ICSSC,Tar18}, and despite the fixed allocation of power and spectral resources, those beams with low traffic demand can serve some users in neighbour beams under more traffic load. 
 With more advanced and flexible payloads, the margin for improvement grows, within the technological constraints of the particular architecture\footnote{
 For example, in conventional payloads, different beams are connected to the same High Power Amplifiers (HPA), thus imposing a group of beams power constraint \cite{cocco2018}.}. 
 
This work  shows how the smart mapping of users and beams provides additional flexibility that can complement that already provided by flexible payloads, all under a common user-centric framework\footnote{Some preliminary results, for a simplified resource assignment scheme,  were presented in \cite{Ram21}.} where the end goal is the matching of the requested user traffic and offered user rates.  A demand-driven radio resource assignment will determine the carrier allocation under practical constraints for power and bandwidth; in the specific case under study,   each HPA will be shared by a pair of beams. Additionally, 
adjacent beams will not be allowed to reuse the same bandwidth \cite{cocco2018,paris2019}, thus neglecting the  co-channel interference\footnote{A non-uniform power allocation across beams can increase the co-channel interference floor at some points; by neglecting it, the results for  flexible power allocation can be slightly optimistic. As opposed to this, the co-channel interference will be strictly monitored when mapping users and beams.}. It should be noted that the proposed methodology is agnostic to the satellite antenna technology. The beam shape, size, and position are assumed to be fixed in this study, though the proposed ideas can be complemented with adaptive beamforming solutions where the distance between beams centers can be adjusted \cite{2021ICSSC},  or both beam size and placement are modified to balance the traffic load \cite{Ang2021}.


The complete user-centric resource allocation problem is non-convex and complex to solve. In consequence, a practical  two-step optimization process will be  proposed: power, bandwidth and users are allocated to beams in a first step, whereas the second step allocates carriers to users within each beam.   In the literature, different  algorithms can be found to tackle the beam resource allocation problem. In particular, a simulated annealing method is employed in \cite{cocco2018}, whereas a genetic algorithm is employed in \cite{paris2019}. Machine learning is applied   to improve the convergence of beam hopping pattern solutions in  \cite{lei2020beam,Xu2020}, and a comparison of different dynamic resource allocation algorithms under realistic operational assumptions is performed in \cite{Garau2020}. Unlike the presented solutions in \cite{cocco2018},\cite{paris2019},\cite{lei2020beam} and \cite{Xu2020}, the first step of the proposed algorithm, operating at the beam level, is driven by  a convex optimization problem that simplifies the beam resource assignment. A convex approach is also found  in \cite{Wang2014,lagunas2020flexible_jour}, under the assumption of a single super-user per beam, though, with a rigid beam-user mapping. For benchmarking purposes, a genetic algorithm will be employed for the joint power and frequency allocation, non-convex problem as illustrated in  \cite{paris2019}. In short, the main contributions of this work are summarized as follows:
\begin{itemize}
    \item A user-centric resource allocation framework is presented that embeds different payload possibilities when allocating resources to beams, accommodating also a flexible beam-user mapping.
    \item A two-step optimization process is proposed as a practical implementation for the complex resource allocation problem, making use of convex optimization and mixed binary quadratic programs (MBQP). 
    \item The benefits of the flexible beam-user mapping in the resource allocation are analyzed by comparing the performance against different flexible payload configurations. The traffic profile across beams is shaped by the Dirichlet distribution, which can be conveniently parameterized to account for different scenarios, and makes simulations  easily reproducible.
\end{itemize}

The rest of the paper is organized as follows.  First, the system model is described in Section \ref{sec:system}. Next, the  resource allocation problem is presented together with the proposed two step optimization framework in Section \ref{sec:Res_All_Problem}. In the subsequent Sections \ref{sec:Beam_fixed} and \ref{sec:Beam_flex},  designs for the resource allocation at beam level are presented for different degrees of freedom. The performance of the explored techniques is presented and compared in Section \ref{sec:Perf}. Finally, some conclusions are given in \ref{sec:Con}.

For the sake of reproducibility of the research presented in this paper,
all the results can be generated with  the Matlab code that is available at  
\url{https://github.com/tomramzp/ResourceAssignmentOneDimesion}. \\

 \noindent \textit{Notation:} Boldface letters denote vectors.  $\ex{\cdot}$ is the expected value operator. For reference purposes, a parameter list is included in Table \ref{tab:notation}.


\section{System Model}\label{sec:system}

\begin{figure}[!htb]
		\centering
		\begin{tabular}{c}
			\includegraphics[width=0.5\textwidth]{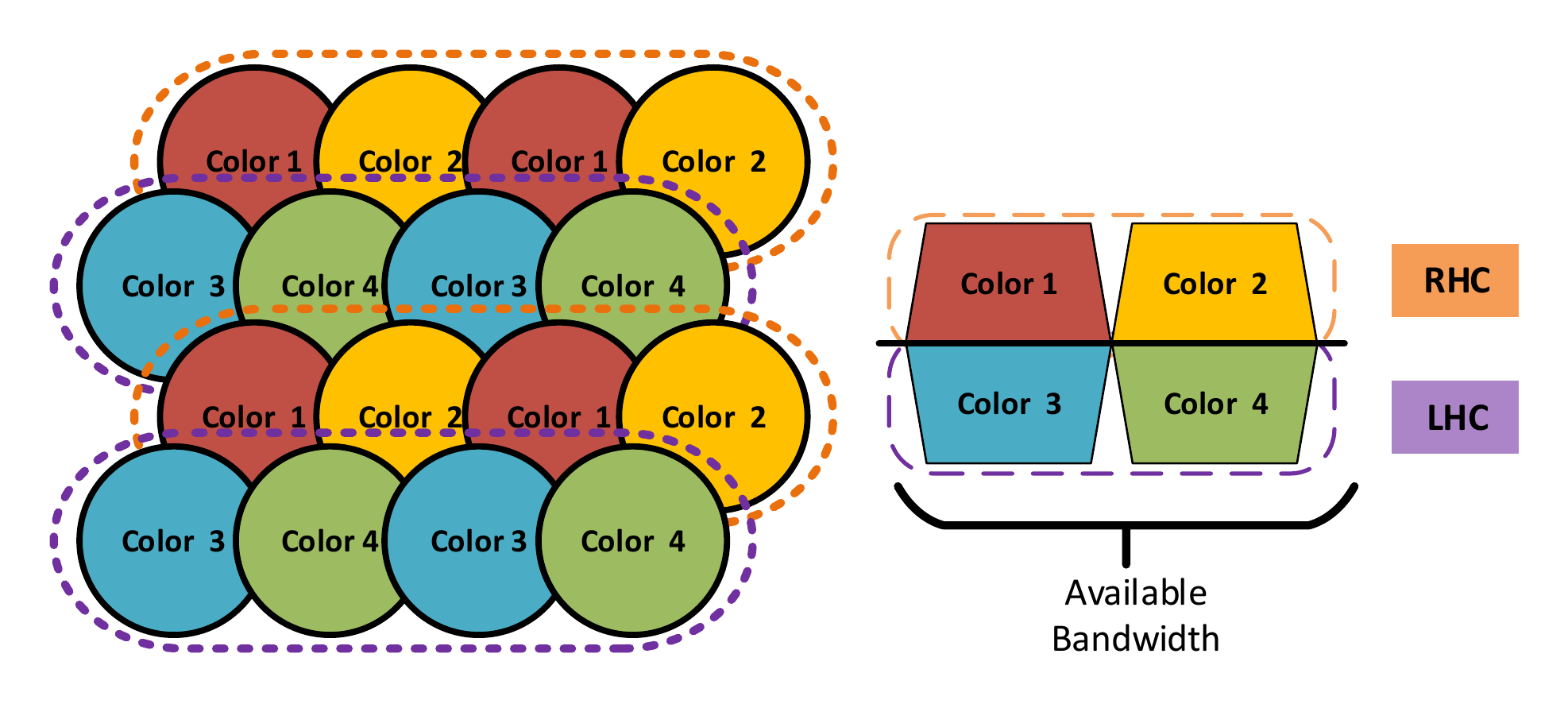}   \\  (a) \\  	\includegraphics[width=0.5\textwidth]{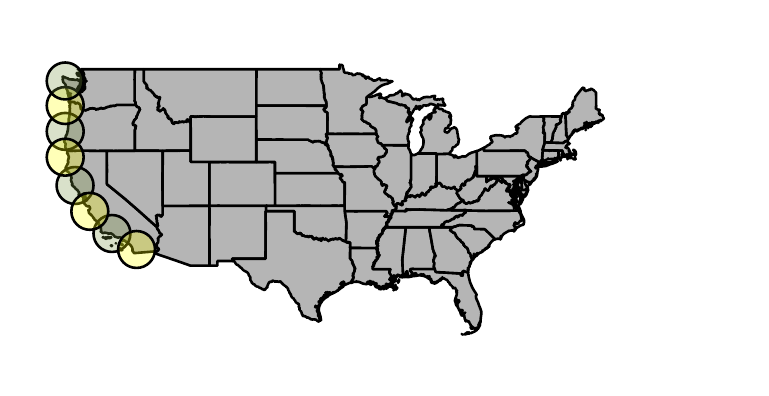}     \\ 	 (b) 
		\end{tabular}
		\caption{ Examples of  one-dimensional satellite coverage.
			(a) Four color reuse scheme. Each row of beams operate under the same polarization.  (b) Two-color scheme over a coastal area. Partial Viasat-1 beam footprint \cite{paris2019}.}
		\label{fig:example_scen}
	\end{figure}

 	A geostationary (GEO) satellite  that illuminates, as part of its coverage,  a row of $K$ consecutive  beams is considered, with a fixed antenna radiation pattern, as  presented in Fig. \ref{fig:example_scen}. This is a very relevant scenario in practical deployments, for example under four color  reuse schemes, for which each color corresponds to a specific half of the available spectrum and polarization; in these cases, we address the rows of beams making  use of the same polarization. Two color reuse schemes, alternating colors in geographically consecutive beams,  are also of interest.
 	 The number of users to be served at a given time instant in the corresponding row of beams is $N$. 
 	The relative location of users and beams is characterized by  ${\mathcal N}(b)$, which  denotes   the set of users on the footprint of  beam $b$, with magnitude  $|{\mathcal N}(b)|$ such that $N = \sum_{b=1}^K |{\mathcal N}(b)|$. In this work, we will use interchangeably beam footprint and  cell to denote the  area on the Earth getting the highest radiation gain  from a  given beam. The set of users served by the beam $b$ will be denoted by 
 	 ${\mathcal S}(b)$. 
	
For the payload, we assume that a given HPA amplifies the whole bandwidth $W^{total}$, so that the pair $j$ of two consecutive  beams, with indexes $A(j)=\{2j-1,2j\}, j=1,\ldots,K/2$, is served by the same amplifier. Colors alternate between consecutive beams to reduce co-channel interference, with no bandwidth overlapping. The available bandwidth $W^{total}$ is split into $2M$  carriers with a fixed carrier bandwidth $W^{\sim}= W^{total}/2M$. Thus, for $M_b$ carriers assigned to the beam $b$, the corresponding bandwidth is  $M_b W^{\sim}$.   We will work interchangeably with the continuous bandwidth $W_b$ and the discrete number of carriers per beam $M_b$, depending on the specific context. If an advanced payload with bandwidth flexibility is assumed, the bandwidth can be split unevenly among the beams sharing the same HPA. Alternatively, if a uniform allocation of carriers is assumed, then  $M_b=M \; \forall \; b$. For both scenarios, and as mentioned above, co-channel interference is neglected by preventing the same carrier from being used in two adjacent beams,  with $  M_a + M_{b}      \leq  2M   \; \forall \; (a,b) \; \text{adjacent beams}$.
	
	As to the carrier power, let $p_b^{\sim}$ be  the power of the carriers that are assigned to the beam $b$. We assume an equal power allocation for all the carriers under the same HPA, so  that $p_a^{\sim}=p_b^{\sim}$ for two beams $a$ and $b$ connected to the same HPA. Thus, we avoid 
	the so-called capture effect in satellite systems with Traveling-Wave Tube Amplifiers (TWTA) when carriers with different levels are fed to a given  power amplifier \cite{cocco2018,maral2020satellite}. Note that  no additional impairments are assumed for the power amplification, which is considered to operate linearly. If $P_j^{PA}$ denotes the output power of the amplifier $j$ that serves the beams $a$ and $b$, then the carrier power is given by\footnote{If $ |{\mathcal S}(b)| < M_b$, not all carriers are employed to serve the users. It is assumed that the power of non-active carriers cannot be allocated to active carriers.} 
 \begin{equation}
       p_a^{\sim}=p_b^{\sim}= \frac{P_j^{PA}}{ M_a + M_b }, 
\end{equation}
and one has that the  power assigned to the beam $b$ is 
 \begin{equation}
       P_b=p_b^{\sim}M_b. 
\end{equation}
 The use of Multi-Port Amplifiers (MPA) furnishes the payload with  a flexible allocation of the power, which can be shared by the different HPAs under the same MPA. For simplicity, we will consider an overall power budget of  $P^{total}$, and a maximum power cap per HPA $P^{max}$ such that $P_j^{PA} \leq P^{max} \; \forall \; j$, although additional power constraints per different groups of HPAs could be considered.
	Alternatively, if a rigid power allocation is enforced, then all carriers are uniformly sent from the satellite with the same power as
 \begin{equation}\label{eq:power_car_uni}
      p_{uni}^{\sim}= \frac{P^{total}}{KM}. 
\end{equation}
For the characterization of the radiation pattern, we resort to the Bessel function model \cite{Bessel},  which facilitates the reproducibility of this study. Under this radiation pattern model, the channel gain from the $b$-th beam towards an $n$-th user within the central beam can be expressed as 
\begin{equation}
    g_b(n)= G_{max} \left(   \frac{J_1(u)}{2\cdot u}  +36 \cdot \frac{J_3(u)}{u^3} \right)^2,
\end{equation}
with $u\approx 2.07123 \cdot d_b(n)/R$, and $d_b(n)$  the distance between the $b$-th beam boresigth and  the $n$-th user location. 
Under this model, the end-to-end channel response (including the antenna gain) of the $n$th user with respect to beam $b \in \{1,\ldots,K\}$ is denoted as  $h_b(n) $, and the channel magnitude is obtained as 
\begin{equation}
|h_b(n)| = \frac{\sqrt{G_R  g_b(n)}}{ \sqrt{k_{\text{B}} T W^{\sim} }  4\pi D/\lambda}
\label{eq:hbn}\end{equation}
where $G_R$ refers to the receive antenna gain, $\lambda$ is the carrier wavelength, and $k_{\text{B}}$, $T$ denote the Boltzmann constant  and the clear sky noise temperature, respectively. Finally, the placement of the beams in the row layout is such that a distance $2R$ separates adjacent beams, for a beam radius  $R$ in the Bessel model.

 \begin{table}[!htb]
	\centering
		\begin{tabular}{|l|l|l|}
			\hline
			\rowcolor[HTML]{C0C0C0} 
			\textbf{Parameter} & \textbf{Description}\\ \hline
			$K$ & Number of beams  \\ \hline
			${\mathcal N}(b)$ & Set of users on the footprint of  beam $b$ \\\hline
			${\mathcal S}(b)$ & Set of selected users to be served by the beam $b$  \\  \hline
			$N$ & Total number of users requesting traffic  \\ \hline
			$\omega_{c,n}$ & \makecell[l]{Time fraction of  carrier $c$ allocated to the  \\ user $n$ in the beam $b$, $n\in {\mathcal S}(b)$ }\\ \hline
			\bm $\Omega_b$  &  Vector that collects the carrier filling at beam $b$ \\ \hline
			\bm $\Omega$   &  Vector that collects the  overall carrier filling   \\ \hline
			$u_{c,n}$ & \makecell[l]{Binary variable that denotes if the user $n$ is active in \\ the carrier $c$ of the beam $b$, $n\in {\mathcal S}(b)$ }\\ \hline
			$\bm U_b$    &   Vector that collects the carrier assignment for beam $b$ \\ \hline
			$\bm U$    &   Vector that collects the overall  carrier assignment \\ \hline
			$W^{total}$ & Total bandwidth   \\ \hline
			$W_b$ & Bandwidth allocated to the beam $b$  \\ \hline
			$\bm W$ & Vector that collects the beam bandwidth allocation \\ \hline
			$M_b$   & Number of carriers allocated to beam $b$  \\ \hline
			$W^{\sim} $ & Carrier bandwidth  \\ \hline
			$P_b^{\sim}$ & Power per carrier at beam $b$    \\ \hline
			$P_b$ & Power allocated to beam $b$    \\ \hline
			$P_j^{PA}$ & Power allocated to the $j$th HPA  \\ \hline
			$P^{max}$ & Maximum power per HPA  \\ \hline
			$P^{total}$ & Total power  \\ \hline
			$\bm P$ & Vector that collects the beams power allocation \\ \hline
			$h_b(n)$ & End-to-end channel of user $n$ with respect to beam $b$  \\ \hline
			$R^{req}(n)$ & Requested rate by the $n$th user   \\ \hline
			$R^{off}(n)$ & Offered rate to the   $n$th user   \\ \hline
			$\zeta_{max}$ &  \makecell[l]{Maximum number of carriers that can be\\processed simultaneously by the receivers} \\ \hline
	        $\mathsf w_{n,b}$	&  Bandwidth portion allocated to the $n$th user by beam $b$	\\ \hline
	       $p_{n,b}$	&  Power portion allocated to the $n$th user by beam $b$	\\ \hline
		   	\end{tabular}
	\caption{Notation summary.} 
	 \label{tab:notation}
\end{table}

\section{Resource Assignment}\label{sec:Res_All_Problem}

The purpose of the system under study is to match the offered rates $R^{off}(n)$ to the user  demanded rates $R^{req}(n)$.  This approach, which  is commonly found in the literature under different constraints and variants, see, e.g., \cite{cocco2018,lei2020beam,choi2005,Wang2014,aravanis2015}, will be elaborated for the different configurations under a common framework, with the aim of developing simple optimization problems. 

\subsection{Optimization criteria}

As driving metric for the allocation of resources, we use the  quadratic unmet demand, since it offers a good balance between the satisfaction of the requested demand and user fairness \cite{choi2005,Wang2014}.  
The quadratic unmet objective function reads as  
\begin{equation}\label{eq:unmet}
 \sum_{n=1}^N (R^{req}(n) - R^{off}(n))^2,  
\end{equation}
that, for minimization purposes, can be alternatively expressed, by leaving out the fixed  requested traffic term $R^{req}(n)$, as
\begin{equation}\label{eq:unmet2}
\mathcal U = \sum_{n=1}^N (R^{off}(n))^2 - 2 \sum_{n=1}^N R^{req}(n) R^{off}(n).
\end{equation}

\subsection{Problem Formulation}

\begin{figure}[!htb]
		\centering
			\includegraphics[width=0.25\textwidth]{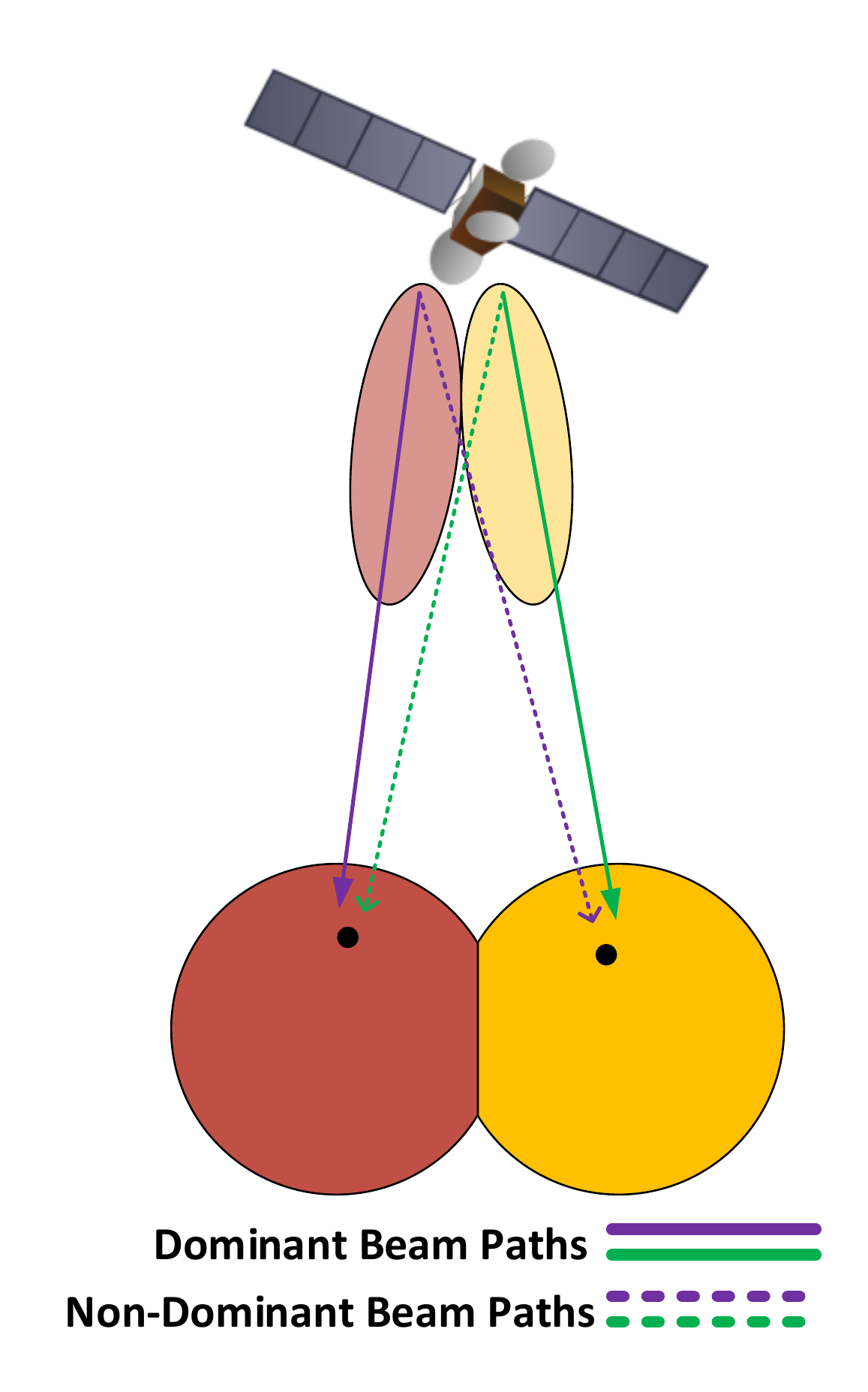}
		\caption{ Different communication paths for flexible beam-user mapping.}
		\label{fig:paths}
	\end{figure}

Initially, we consider a fully flexible architecture, or equivalently, the least restrictive approach, with  
 a satellite payload that can optimize the 
 power and bandwidth allocated to the different users. In addition, the gateway has the capability of freely managing the beam-user mapping, so that users are not necessarily served by their dominant beams. As it is sketched in Fig \ref{fig:paths}, the flexible mapping allows for the potential pairing of a user with a non-dominant beam, such that the radiation intensity is still significant, which precludes all beams except the adjacent one for typical directive radiation diagrams.  
 
 The offered rate to the users served by beam $b$ can be expressed as 
\begin{align}\label{eq:R_off}
R^{off}(n)  =  \sum_{c=1}^{M_b} \omega_{c,n} W^{\sim}\log_2 \left( 1+ \frac{ p_b^{\sim} |h_b(n)|^2} {N_o W^{\sim}} \right) , \, n \in {\mathcal S}(b) 
\end{align}
where $\omega_{c,n}$ parameterizes the fraction of time  that user $n$ uses the baseband frame of carrier $c$ of the beam $b$. Furthermore, with  the variable $u_{c,n}$ denoting if  the user $n$ is active in the carrier $c$ of the beam $b$,  a carrier aggregation constraint \cite{lagunas2020CA} can be set as   
\begin{equation}\label{eq:CA_constraint}
 \sum_{c=1}^{M_b} u_{c,n}      \leq  \zeta_{max} \;    ,    \forall \, n \in  {\mathcal S}(b) 
\end{equation}
where $\zeta_{max}$ is the maximum number of carriers that can be processed simultaneously by the receivers. In this paper, conventional user terminals are assumed, so that $\zeta_{max}=1$ for the single-carrier operation of the receive terminals.

Under this initial and most flexible approach,  the user demand can be  matched through the allocation of bandwidth (in continuous or discrete form, $W_b$ or $M_b$, respectively) and power $P_b$ to the different beams, complemented by a non-rigid mapping between users and beams.  Let $\bm P = [P_1,\ldots,P_{K}] \in   \mathbb{R}^{K}$ and $\bm W = [W_1,\ldots,W_K] \in   \mathbb{R}^{K}$  be the vectors that collect the power and bandwidth assignment per beam, respectively. 
The vectors $\bm \Omega_b = [\omega_{1,1},\ldots,\omega_{M_b,1},\,\omega_{1,2},\ldots,\omega_{M_b,|{\mathcal S}(b)|}] \in   \mathbb{R}^{M_b \cdot |{\mathcal S}(b)| }$ and $\bm U_b = [u_{1,1},\ldots,u_{M_b,1},\,u_{1,2},\ldots,u_{M_b,|{\mathcal S}(b)|}] \in   \mathbb{R}^{M_b \cdot |{\mathcal S}(b)| }$,   collect the carrier allocation and  assignment for  beam $b$ in 
\eqref{eq:R_off} and \eqref{eq:CA_constraint}, respectively, which are grouped in  $\bm \Omega = [\bm \Omega_1,\ldots,\bm \Omega_K] $ and $\bm U = [\bm U_1,\ldots,\bm U_K]$, both vectors of dimension $\sum_{b=1}^K M_b |{\mathcal S}(b)|$.

In order to determine the  bandwidth, power and mapping settings, the minimization of the unmet demand is formulated, jointly  embracing the  power and bandwidth allocation across the beams,  the beam-user mapping, and  the user-carrier mapping within each beam: 
	\begin{equation} \label{eq:Gen_Opt}
	\begin{array}{ll}
	 \displaystyle \min_{\bm W,\bm P,\bm \Omega, \bm U,{\mathcal S}(b) }  \quad  &  \mathcal U    \\
	\mbox{subject to} & \\ 
		 \hphantom{ \mbox{subject to}} \text{C1:} &   \sum^{K/2}_{j=1} P_j^{PA}     \leq  P^{total}\\
	 	 \hphantom{ \mbox{subject to}}   \text{C2:} &    \sum_{b \in A(j)} P_b  \leq P^{max}  ,\; j=1,\ldots,K/2      \\
	 \hphantom{ \mbox{subject to}} \text{C3:} &  W_b + W_{b+1}      \leq  W^{total}    \quad \forall \; b \\
	 \hphantom{ \mbox{subject to}}   &  M_b = \floor{\frac{W_b}{W^{\sim}}} \;  \forall \; b  \\
	 \hphantom{ \mbox{subject to}}  &    0  \preceq \bm  W       \preceq    W^{total}  \\
	 \hphantom{ \mbox{subject to}}  &   \sum_{n \in {\mathcal S}(b)}   \omega_{c,n}       \leq 1  \;  \forall \; b ,\; c=1,\ldots,M_b   \\
	 \hphantom{ \mbox{subject to}}  \text{C4:} &    \sum_{c=1}^{M_b} u_{c,n}      \leq  1 \;   \forall  \; n  \in {\mathcal S}(b) \\
	 \hphantom{ \mbox{subject to}}  &    u_{c,n}   \geq  \omega_{c,n}    \\
	 \hphantom{ \mbox{subject to}}  &   0 \leq\omega_{c,n}   \leq 1    \\
	 \hphantom{ \mbox{subject to}}  &   u_{c,n}   \in \{0,\; 1\} \; \forall \; b,c,n \in {\mathcal S}(b) \\
	\end{array}
	\end{equation}
where the constraint  C1 is set to satisfy the  overall power constraint, C2 limits the amount of power per HPA, and  C3 avoids the use of the same portion of spectrum in adjacent beams to reduce the co-channel interference. Finally, C4 constraints the terminals to operate in single carrier mode.\\
 
The formulated optimization problem is non-convex due to the interplay of discrete and continuous variables, and difficult to solve. In order to make the problem more easily tractable, we will follow a  two-step approach, first across beams, prior to the resource allocation inside each beam. This strategy will be followed for the different architectures in the next sections, all subsumed by the general optimization problem  in \eqref{eq:Gen_Opt}\footnote{The proposed two-level process could also be applied to other  performance indicators if a convex formulation of the objective function can be written.}:
\begin{itemize}
\item \textit{First step (beam level)}: Following a  selected strategy, the resources at beam level are optimized in terms of power $P_b$, bandwidth $W_b$,  and/or user assignment ${\mathcal S}(b)$. 

\item  \textit{Second step (intra beam):} After the beam resources are allocated, the carrier power per beam $p_b^\sim$ and the discrete number of carriers per beam $M_b$ are obtained, with the carrier assignment  performed independently for each beam. The following optimization problem solves the user-carrier assignment within the beam $b$:
	\begin{equation}\label{eq:carrier-mapping}
	\begin{array}{ll}
	 \displaystyle \min_{\bm \Omega , \bm U }  & \mathcal U_b \triangleq  \displaystyle \sum_{n \in {\mathcal S}(b)} R^{off}(n))^2  -   \\ & \hphantom{ R^{off}(n))^2} 2 R^{req}(n)R^{off}(n)  \\
	\mbox{subject to}  	  &  \sum_{n \in {\mathcal S}(b)}   \omega_{c,n}       \leq 1, \forall \; c=1,\ldots,M_b \\
	\hphantom{ \mbox{subject to}}  & \omega_{c,n} \geq 0, \forall \; c,n \\
	\hphantom{ \mbox{subject to}} &  \sum_{c=1}^{M_b} u_{c,n}     \leq  1  \quad   \forall \; n \in {\mathcal S}(b)\\ 
	\hphantom{ \mbox{subject to}}&   u_{c,n}   \geq  \omega_{c,n}    \\
	\hphantom{ \mbox{subject to}}  &   u_{c,n}   \in \{0,\; 1\} , \forall \; c,n \in {\mathcal S}(b). \\ 
	\end{array}
	\end{equation}
	 This problem is an  MBQP that can be solved by conventional methods such as branch and bound and/or the outer approximation approach \cite{Branch,Outer}.

\end{itemize}
 Our main goal is the derivation of optimization problems for the different cases, which are amenable for the practical implementation of the corresponding solutions. In the following sections we address the beam level optimization for different degrees of flexibility, initially for a fixed beam-user mapping, prior to a non-rigid pairing of beams and users.

 \section{Beam level optimization for fixed beam-user mapping}\label{sec:Beam_fixed}
 
 We start by exploring solutions that allow the flexible allocation of power and/or bandwidth across beams for a conventional user assignment to their dominant beams. Under this rigid mapping, we have that ${\mathcal S}(b) = {\mathcal N}(b)$, i.e., 
 the set ${\mathcal S}(b)$ of users served by the beam $b$ coincides with the set of users ${\mathcal N}(b)$ within  its cell. Now,  the quadratic unmet demand in  \eqref{eq:unmet2} can be 
alternatively expressed after grouping users at the different beam  cells: 
\begin{equation}\label{eq:unmet_alt}
\mathcal U  =\sum_{b=1}^K  \sum_{n \in {\mathcal N}(b)} \left[ R^{req}(n)R^{off}(n) - \frac{1}{2}  (R^{off}(n))^2 \right].
\end{equation}
In order  to decouple the optimization problem in two steps, we assume that all users are offered the same rate by beam $b$: 
		\begin{equation}
   	R^{off}(n) = \frac{1}{|{\mathcal N}(b)|}  R_{beam}^{off}(b)    , \; \forall \; n \in {\mathcal N}(b)   
	\end{equation}
	for an aggregated beam rate $R_{beam}^{off}(b)$, approximated by means of an effective SNR ${\snr}_b$, as
    \begin{equation}\label{eq:R_off_3}
    R^{off}_{beam}(b)   \approx W_b \log_2(1+ x_b \cdot  {\snr}_b)  
    \end{equation}		
   where $0 \leq x_b \leq 1$ is the fraction of total power allocated to beam $b$ such that $P_b= x_b P^{total}$. The effective SNR ${\snr}_b$ collects in a unique parameter the offered rate to the users across the beam $b$ with different channel qualities. This effective SNR ${\snr}_b$ is defined from the following equality:
	\begin{multline}\label{eq:effective_snr_def}
	\log_2(1+ x_b \cdot  {\snr}_b) \\  = \frac{1}{|{\mathcal N}(b)|}  \sum_{n \in {\mathcal N}(b)} \log_2\left(1+ \frac{x_b P^{total} |h_b(n)|^2}{N_o W_b} \right),
	\end{multline}
	and can be approximated for moderate to high SNR values, as 
 \begin{equation}\label{eq:effective_snr_gen}
	{\snr}_b \approx \left( \prod_{n \in {\mathcal N}(b)} \frac{P^{total} |h_b(n)|^2}{N_o W_{b}}   \right)^{ \frac{1}{|{\mathcal N}(b)|}}.
	\end{equation}
	With this,  the unmet function in \eqref{eq:unmet_alt} can be written at the beam level as 
	\begin{equation}\label{eq:unmet_alt_approx}
\mathcal U  =	\sum_{b=1}^K \frac{1}{|{\mathcal N}(b)|} R_{beam}^{req}(b)R_{beam}^{off}(b) - \frac{1}{2} \frac{1}{|{\mathcal N}(b)|} (R_{beam}^{off}(b))^2
	\end{equation}
     where
	\begin{align}
      R_{beam}^{req}(b) & =   \sum_{n \in {\mathcal N}(b)}  R^{req}(n).   
		\end{align}

This metric is used in the next subsections  for devising different criteria for the beam resource allocation. The presented designs are collected in Table \ref{tab:tech_no_UM}, where a beam level optimization (first step) is explored for power, bandwidth, and both simultaneously. It has been previously reported in the literature the limited gain from power flexibility on top of bandwidth flexibility, see, e.g., \cite{cocco2018} or \cite{paris2019},  which can be theoretically traced back  to early results in \cite{cioffi2000}. In any case, for benchmarking purposes, we will use population based metaheuristics, in particular, a genetic algorithm, to address the joint power and bandwidth allocation problem  under the power and bandwidth constraints defined in Section \ref{sec:system}. 

\begin{table*}[!htb]
	\centering
	\begin{tabular}{|l|c|c|c|}
		\hline
		\rowcolor[HTML]{C0C0C0} 
		\textbf{Technique} & \textbf{Power per carrier} & \textbf{Carriers per beam} & \textbf{Optimization problem} \\ \hline
		Beam Power Allocation &  Variable & Fixed  & convex QP
		\\\hline 
		Beam Bandwidth Allocation &  Fixed  & Variable &  convex QP \\ \hline
		Beam Power \& Bandwidth Allocation & Variable & Variable &  non-convex QP \\ \hline 
	\end{tabular}
	\caption{Beam-level optimization (first step) for the fixed beam-user  mapping solutions.}\label{tab:tech_no_UM}
\end{table*}

\subsection{Flexible Beam Power Allocation}\label{sec:flexP}
First we consider a uniform bandwidth sharing for all beams, with  flexible power allocation.
The two colors reuse scheme is such that $M$ carriers are available per beam, for a total of $2M$ carriers, whereas the fractions of total power per beam values $x_b,b=1,\ldots,K$, collected in $\bm x$, are the optimization parameters. Following the presented two-step process, the power allocation across beams is obtained in the first step after maximizing \eqref{eq:unmet_alt_approx}:
\begin{equation}\label{eq:opt-power2}
\begin{array}{ll}
\mbox{max} & g(\bm x) =  \displaystyle \sum_{b=1}^K \frac{1}{|{\mathcal N}(b)|} R_{beam}^{req}(b)R_{beam}^{off}(b) - \\  
 	 \hphantom{\mbox{max}}   &   \hphantom{g(\bm P) =}    \frac{1}{2} \frac{1}{|{\mathcal N}(b)|} (R_{beam}^{off}(b))^2 \\  \\ 
 	 \mbox{subject to}    &    \displaystyle \sum_{b \in A(j)} x_b  \leq \frac{P^{max}}{P^{total}}  ,\; j=1,\ldots,K/2      \\   
 	 \hphantom{ \mbox{subject to}}  &   \displaystyle \sum_{b=1}^K x_b      \leq  1  ,\; j=1,\ldots,K/2 \\
\end{array}
\end{equation}
with
\begin{equation}
R_{beam}^{off}(b)  =  \sum_{n \in {\mathcal N}(b)}  R^{off}(n)  = \frac{W^{total}}{2} \log_2\left(1+ x_b \cdot {\snr}_b \right).
\end{equation}
Note that the effective SNR ${\snr}_b$ in  \eqref{eq:effective_snr_gen} is defined for the same bandwidth  $W_b=W^{total}/2$ across beams. 
This problem turns out to be convex, as proven in \ref{sec:appA}.

\subsection{Flexible Beam Bandwidth Allocation}\label{sec:flexBW}

For the same two colors reuse scheme as above, we explore next the flexible allocation of carriers to beams, in an effort to allocate more frequency resources to more demanded beams, while still keeping the co-channel interference at low levels. The power per carrier, uniform for all beams,  was  given by  \eqref{eq:power_car_uni}, so that only the bandwidth flexibility is exploited  by allowing different numbers of carriers per beam. Note that this results in a variable beam power allocation that is proportional to the bandwidth\footnote{Due to the time/frequency duality, both flexible bandwidth and beam-hopping schemes will provide a similar result, so no specific mention will be made to the application of beam-hopping.}.

The optimization problem \eqref{eq:opt-power2} is rephrased for the  bandwidth allocation per beam $\bm W$ as 
\begin{equation}\label{eq:opt-bw}
\begin{array}{ll}
\mbox{max} & g(\bm W) = \sum_{b=1}^K \frac{1}{|{\mathcal N}(b)|} R^{req}_{beam}(b)R^{off}_{beam}(b) - \\
 	 \hphantom{\mbox{max}}   &   \hphantom{g(\bm P) =}  \frac{1}{2} \frac{1}{|{\mathcal N}(b)|} (R^{off}_{beam}(b))^2 \\ \\
  \mbox{subject to}  &   W_a + W_b \leq W^{total}, \forall \; (a,b) \; \text{adjacent beams}.\\       
\end{array}
\end{equation}
With the power per beam proportional to the corresponding allocated bandwidth, the offered rate at beam $b$ is expressed as 
\begin{equation}
R^{off}_{beam}(b)  = W_b \log_2\left(1+ \frac{1}{K}{\snr}_b \right)
\end{equation}
with ${\snr}_b$ in \eqref{eq:effective_snr_gen}.  With this, \eqref{eq:opt-bw} turns out to be  a convex problem which provides the bandwidth allocation vector $\bm W$. 

The allocated bandwidth $W_b$ needs to be formulated  as a discrete number $M_b$ for the intra-beam carrier-user mapping.  Thus, the number of carriers per beam is obtained as
\begin{equation}
M_b = \frac{W_b}{W^{total}/2} M.
\end{equation}
As computed, $M_b$ will not be in general integer, so that it must be rounded in such a way that the same carrier is not used in adjacent beams.

\subsection{Flexible Power and Bandwidth Allocation}\label{sec:GA}

For benchmarking purposes it is desirable to evaluate the performance of a  potential scheme able to allocate both  bandwidth and power across beams, together with the intra-beam sharing of resources. This general problem, as stated in  \eqref{eq:Gen_Opt}, is a hard one, as described, for example, in \cite{paris2019}. Therein, a genetic algorithm is used, motivated by encouraging results in previous works such as \cite{aravanis2015}. The details of the genetic algorithm which have been implemented for this work can be found in  \ref{sec:appB}.

\section{Beam level optimization for flexible beam-user mapping}\label{sec:Beam_flex}

In the previous section,  power and/or  bandwidth were flexibly allocated to the different beams for a fixed assignment between users and beams. Conventionally, users are served by their dominant beams, even though the additional freedom provided by a flexible mapping between users and beams can be exploited to balance the beam load, as illustrated in previous works dealing with the hot-spot scenario \cite{Nader2017}, \cite{2018ICSSC}, \cite{Tar18}. As in the previous first level designs, the goal is to pose a convex problem for the practical implementation of the resource allocation at beam level. Thereby, the framework for the beam level optimization  derived from \eqref{eq:unmet_alt} needs to be modified to accommodate a non-rigid allocation of users to beams.

A flexible power allocation does not lead easily to a tractable problem in combination with the optimization of the beam-user mapping. In addition, the power flexibility was seen to provide limited gain, at least with respect to the corresponding bandwidth allocation across beams \cite{cocco2018}. Therefore, beam-user mapping will be considered to enhance only flexible bandwidth solutions. The allocation of carriers to beams will be relaxed, so that continuous variables will be considered when assigning spectral resources to beams. All in all, the simplifications that we will apply on the general optimization problem \eqref{eq:Gen_Opt} are listed as:
\begin{enumerate}
    \item[i)] The variables denoting the bandwidth allocated to the different beams will be continuous, thus relaxing the discrete nature of carriers. At the end rounding is applied to obtain an integer number of carriers per beam.
    \item[ii)] The power per carrier is not a free parameter, but rather a fixed predefined value, in such a way that those beams with fewer allocated carriers will consume less power.
\end{enumerate}

For benchmarking purposes, a design with fixed bandwidth allocation and flexible mapping is also assumed.
These designs are collected in  Table \ref{tab:tech_UM}  and presented in the following subsections.
  
\begin{table*}[!htb]
	\centering
	\begin{tabular}{|l|c|c|c|}
		\hline
		\rowcolor[HTML]{C0C0C0} 
		\textbf{Technique} & \textbf{Power per carrier} & \textbf{Carriers per beam} & \textbf{Optimization problem} \\ \hline
			Beam-user Mapping &  Fixed  & Fixed &  Convex QP \\ \hline
		Beam-user Mapping \& Flexible Bandwidth Allocation & Fixed & Variable &  Convex QP \\ \hline
	\end{tabular}
	\caption{Beam-level optimization (first step) for the flexible beam-user mapping solutions. }\label{tab:tech_UM}
\end{table*}

\subsection{Joint optimization of bandwidth and beam-user mapping}

By following the relaxation described above, we can alternatively express the optimization problem  \eqref{eq:Gen_Opt} as the following convex Quadratic Programming (QP) problem:

	\begin{equation}\label{eq:relax_QP_FlexMap}
	\begin{array}{ll}
	   \displaystyle \min_{\bm {\mathsf w}}  &    \mathcal U    \\
	\mbox{subject to}  &  R^{off}(n)= \displaystyle \sum_{b=1}^{K}  \mathsf w_{n,b} \log_2 \left( 1+ \frac{ p_{n,b} |h_b(n)|^2} { N_o  \mathsf w_{n,b}} \right)\\
	 \hphantom{ \mbox{subject to}}  & \displaystyle  W_b= \sum_{n=1}^N \mathsf w_{n,b} \\
	 	 \hphantom{ \mbox{subject}} \text{C1:} & W_b + W_{b+1}      \leq  W^{total}    \quad \forall b \\
	   \hphantom{ \mbox{subject}} \text{C2:}  & \sum_{b=1}^K \mathsf w_{n,b} \leq W^{\sim}  \quad \forall n \\
	 	 \hphantom{ \mbox{subject to}}  &   \mathsf w_{n,b} \geq 	  0
	\end{array}
	\end{equation}
	where  $\bm {\mathsf w} = [\mathsf w_{1,1},\ldots,\mathsf w_{N,1},\,\mathsf w_{1,2},\ldots,\mathsf w_{N,K}] \in   \mathbb{R}^{N \cdot K }$, and $\mathsf w_{n,b}$ and $p_{n,b}$ are the portions of the bandwidth and power, respectively, allocated  to the $n$th user by  beam $b$; note that the power  is proportional to the bandwidth as $p_{n,b}= \mathsf w_{n,b}\frac{P^{total}}{K} \frac{2}{ W^{total}}$. Constraint C1 ensures that the bandwidth is not reused in consecutive beams, whereas  constraint C2  limits the  bandwidth that can be allocated to a given user to that of a carrier.

Once the relaxed problem is solved, the bandwidth of the beam $b$ is obtained as $W_b= \sum_{n=1}^N \mathsf w_{n,b}$,   and the procedure detailed in Section \ref{sec:flexBW} can be followed to obtain the number of carriers per beam $M_b$. The beam-user mapping is also extracted from $\mathsf w_{n,b}$, with  the user $n$  assigned to the beam offering the maximum bandwidth as  
$\arg \max_b \mathsf w_{n,b}$. Note that some users could remain unpaired after the beam-level optimization, for example if they demand low relative rates in highly asymmetric traffic settings. 
  Nevertheless, we want to ensure the mapping of every user in the first step and leave the decision of the resource assignment to the subsequent intra-beam processing in the second step. To this end, we assume that every user is assigned to its dominant beam unless the optimization says otherwise.

\subsection{Beam-user mapping with fixed bandwidth}

If the bandwidth allocation is fixed,  the constraint C1 in \eqref{eq:relax_QP_FlexMap} needs to be reformulated, since the set of available carriers per beam is fixed, and such that $W_b \leq M\cdot W^{\sim}$. The outcome of the optimization problem will be the beam-user assignment; this solution can be applied to conventional payloads with fixed allocation of resources, since the radio resource management  is the only  unit required to enforce this additional flexibility. Other system  implications are discussed in the next section.

\section{ System requirements for flexible beam-user mapping}\label{sec:Req_flex}

In conventional satellite systems, a rigid 
beam-mapping is assumed, for which users need to report a single magnitude value to quantify the quality of the link with their respective dominant beam. If flexible beam-user mapping is applied, then the gateway can serve a given user through a non-dominant beam if required. This flexibility can be exploited if the gateway knows the corresponding channel magnitude; thus, user terminals need to report back the magnitude of  their two downlink strongest channels. Note that, under this paradigm, a given user will be served only by its dominant beam or the strongest adjacent beam; in other words, flexible mapping with non-dominant beam assignment can be enforced if the alternative beam radiates a minimum power at the intended location. 

From the above, the implementation at the system level of the flexible user-mapping requires the knowledge of two channel quality magnitude values per user. Since consecutive beams operate in alternate frequency bands, and the simultaneous channel estimations cannot be pursued with standard single-carrier user terminals, some system add-ons are needed, which can be based on existing solutions. The first solution can be based on the handover mechanisms in the standardized mobile satellite system \cite{UMTS}, where a common frequency can be reused in every beam for the simultaneous channel estimation with the application of unique spreading sequences for each beam. It should be remarked that this common frequency is only employed for the system synchronization and management of the user terminals; conventional orthogonal carriers are employed for on-demand data channels. Alternatively, a user position method can be also devised similarly to the handover management in DVB-S2 \cite{GuidDVB,itDVB}. In that case, user terminals report its location along with the channel estimation of its dominant channel, and the gateway could infer the channel strength of the adjacent beams if a accurate model of  the radiation pattern is available. 
In any case, note that a handover operation can be enforced even for a static user, as it is a system level decision accounting for global metrics\footnote{In case of multiple gateways, the routing of the traffic can be performed at higher layers.}. Finally, let us remark that the proposed flexible beam-user mapping is fully compatible with the existing air interface standards, such as DVB-S2X, 
even avoiding the use of the new features of DVB-S2X supporting   intermittent  transmissions as needed for beam hopping, and with a low demand for the channel state information in the feedback channel, as compared to precoding schemes.

\section{Performance evaluation}\label{sec:Perf}

\begin{table*}[!htb]
	\centering 
	\resizebox{0.99\textwidth}{!}{%
		\begin{tabular}{|c|c|c|c|c|c|c|}
			\hline
			\rowcolor[HTML]{C0C0C0} 
			\textbf{Technique} & \textbf{Label} &\textbf{  \makecell{Carrier\\Power} } & \textbf{  \makecell{Carrier\\Bandwidth} } & \textbf{\makecell{Number of carriers \\ per beam} }  & \textbf{User mapping} & \textbf{Optimization} \\ \hline
			\multirow{2}{*}{	Beam Power Allocation} &  \multirow{2}{*}{POW} & \multirow{2}{*}{  Variable  }  & \multirow{2}{*}{ Fixed  }  & \multirow{2}{*}{ Fixed  }  & \multirow{2}{*}{Dominant beam} &  \multirow{2}{*}{ \makecell{Two step: (i) Convex   \\ (ii) MBQP}  }    \\  
			&   &  &  &  & 	&\\ \hline 
			\multirow{2}{*}{ Beam Bandwidth Allocation} &  \multirow{2}{*}{BW} &  \multirow{2}{*}{Fixed  }  & \multirow{2}{*}{ Fixed }  & \multirow{2}{*}{ Variable  }  & \multirow{2}{*}{Dominant beam} &   \multirow{2}{*}{ \makecell{Two step: (i) Convex   \\ (ii) MBQP}  }   \\  
				&   &  &   &  & 	&\\ \hline 
			\multirow{2}{*}{ \makecell{	Beam Power\&Bandwidth Allocation \\ (Benchmark) }} &  \multirow{2}{*}{BW-POW}  &  \multirow{2}{*}{ Variable  }  & \multirow{2}{*}{ Variable  } & \multirow{2}{*}{ Variable  }  & \multirow{2}{*}{Dominant beam} &   \multirow{2}{*}{ \makecell{Two step: (i) Genetic Algorithm  \\ (ii) MBQP}  }    \\  
			&   &   &   &  & &	\\ \hline 
			\multirow{2}{*}{ \makecell{ Beam-user Mapping \\with fixed resources per beam  }  } &  \multirow{2}{*}{MAP}&  \multirow{2}{*}{Fixed }  &  \multirow{2}{*}{Fixed } &  \multirow{2}{*}{Fixed }   & \multirow{2}{*}{Free assignment} &  \multirow{2}{*}{   \makecell{Two step: (i) Convex   \\ (ii) MBQP}  }    \\  
				&   &   &  &  & &	\\ \hline 
			\multirow{2}{*}{ \makecell{ Beam-user Mapping \\  with flexible Bandwidth per Beam}  } &  \multirow{2}{*}{BW-MAP} &   \multirow{2}{*}{Fixed }   & \multirow{2}{*}{Fixed    } &  \multirow{2}{*}{Variable }  & \multirow{2}{*}{Free assignment} &  \multirow{2}{*}{  \makecell{Two step: (i) Convex   \\ (ii) MBQP}  }    \\
			&   &  &  &  & &	\\ \hline 
		\end{tabular}
	}
	\caption{Summary of the considered techniques.} \label{tab:tech_sum}
\end{table*}

In this section, we evaluate the performance of the proposed design with flexible beam-user mapping for  different payload architectures. A comprehensive summary of the explored techniques is presented in  Table \ref{tab:tech_sum}.  These techniques  will be applied to the model in Section \ref{sec:system} with the parameters in Table  \ref{tab:SystemPar}. The solutions of the corresponding optimization problems are obtained with two tools, namely, the  Mosek solver \cite{mosek} and the optimization toolbox CVX \cite{cvx}, for the mixed integer and  the    convex problems, respectively. As detailed later, the traffic demand density per beam will follow  different traffic profiles across beams; 500 Monte-Carlo realizations will be run to account for random traffic variations. 
Note that, as per Table \ref{tab:SystemPar}, only six beams will be considered to keep the  complexity of the simulations relatively low.  Nevertheless, the analyzed methods operate with an arbitrary number of beams,  and numerical results not  shown here  reveal that conclusions hold for larger scale scenarios.

\begin{table}[H]
	\centering
	\begin{tabular}{cc}
		\hline
		Diagram pattern		    &  Bessel modeling   \\  \hline
		Number of beams & 6 \\  \hline
		Maximum antenna gain &  52 dBi \\ \hline
		Average free space losses &  210 dB  \\ \hline
		Average atmospheric losses & 0.4 dB   \\ \hline
		Total available transmission power, $P^{total}$ & 200 W \\ \hline
		Maximum  power per HPA, $P^{max}$ & 133 W \\ \hline
		Total available bandwidth, $W$ & 500 MHz \\ \hline
		Polarization & Single \\ \hline
		Frequency reuse scheme & 2-Color \\ \hline
		Number of carriers per color, $M$ & 4 \\ \hline
		Frequency band [GHz]	& 20 \\  \hline
		\multicolumn{2}{c}{\textbf{ Receiver Parameters}}  \\   \hline 
	    Terminal G/T &  16.25  dB/K \\  \hline
	    Losses due to terminal depointing &  0.5 dB \\  \hline
	\end{tabular}
		\caption{System parameters for the forward link.}	\label{tab:SystemPar}
\end{table}


\subsection{ Considerations on the co-channel interference }\label{sec:Num_CoChaI}

\begin{figure}[!htb]
		\centering
		\begin{tabular}{cc}
			\includegraphics[width=0.3\textwidth]{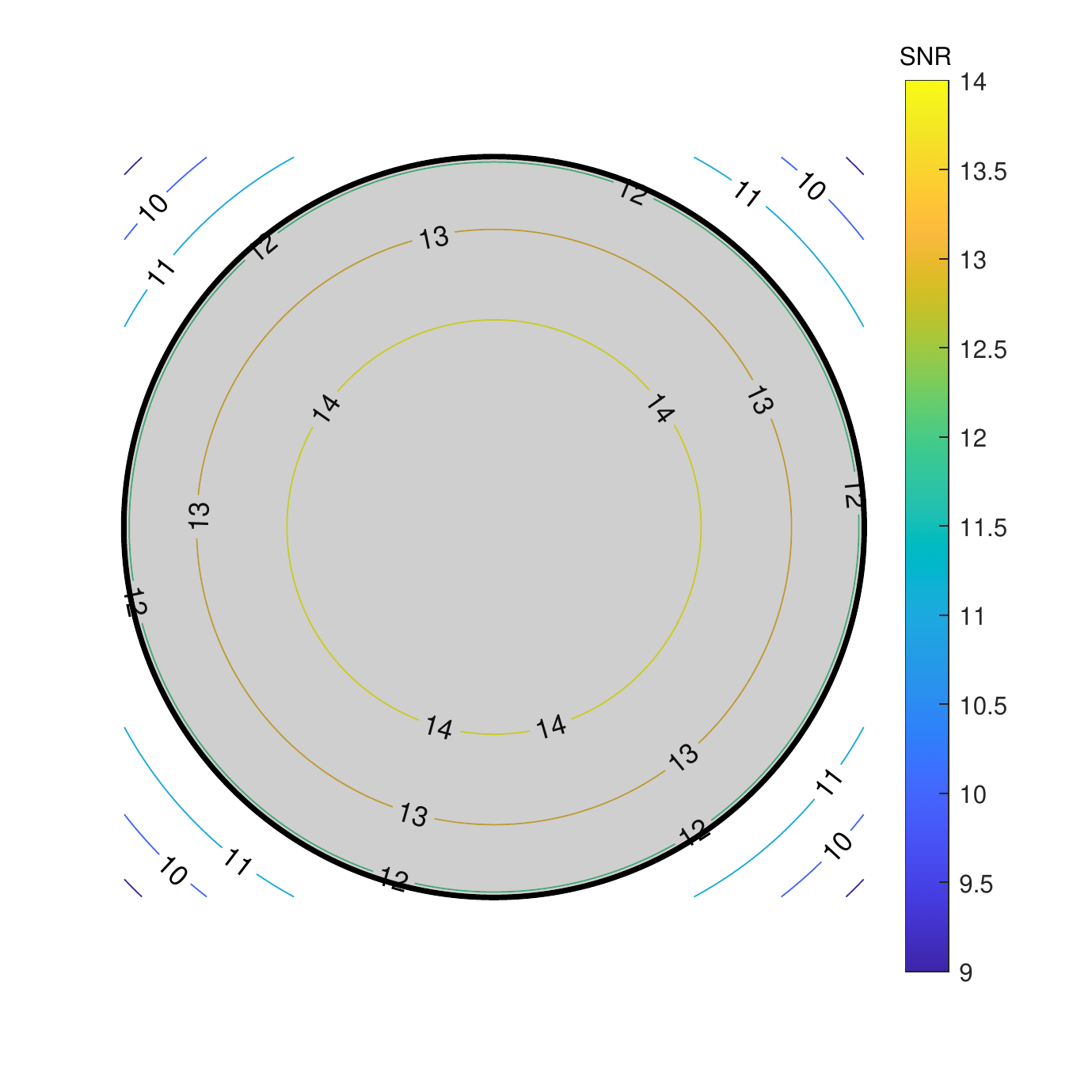}  \\ (a) \\ 	\includegraphics[width=0.3\textwidth]{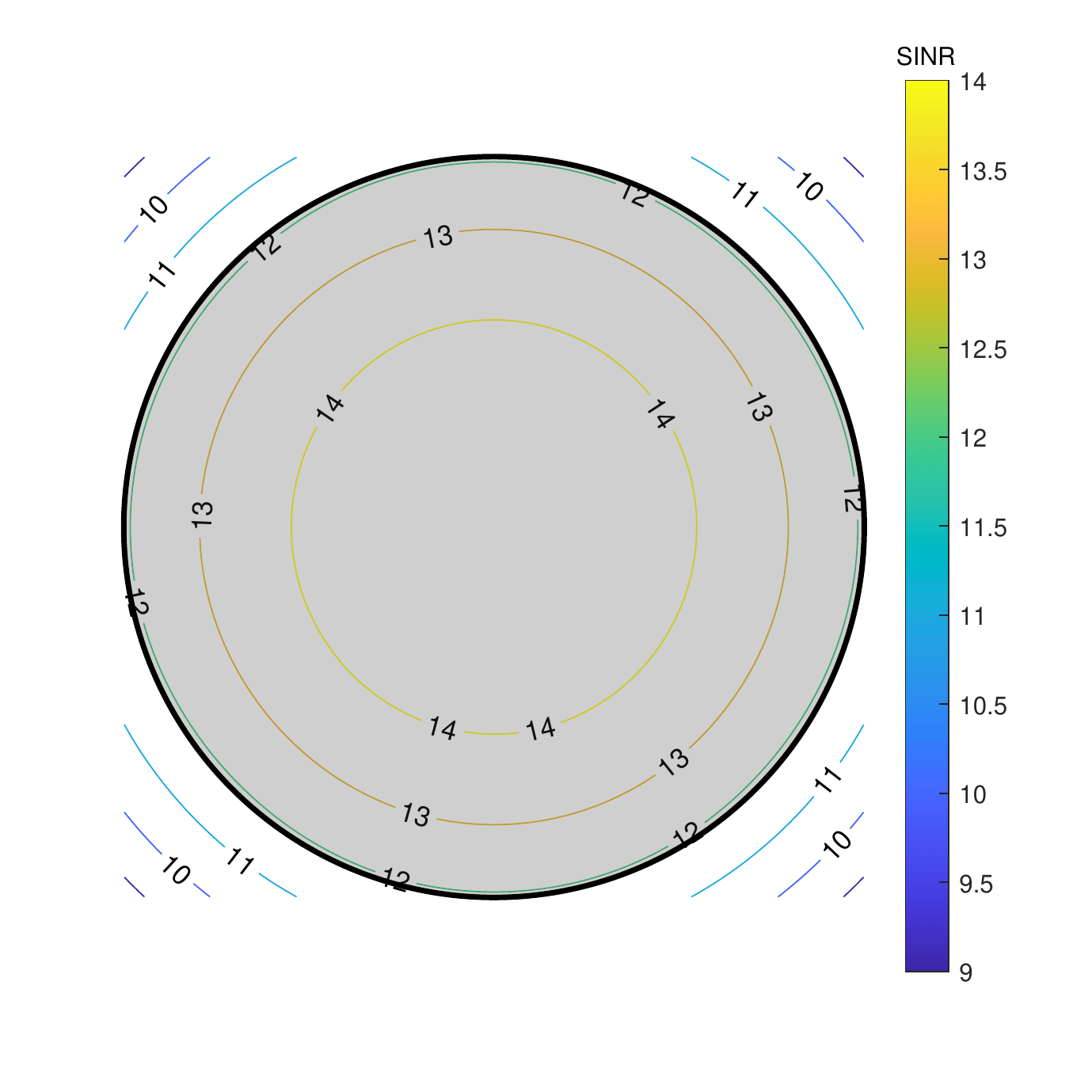}  \\   	 (b) 
		\end{tabular}
		\caption{ Contour map of the signal strength within a beam for uniform resource allocation. The grey area displays the beam footprint at 3 dB. (a) SNR,  (b) SINR.}
		\label{fig:snr_sinr_dom}
	\end{figure}

\begin{figure}[!htb]
		\centering
		\begin{tabular}{cc}
			\includegraphics[width=0.3\textwidth]{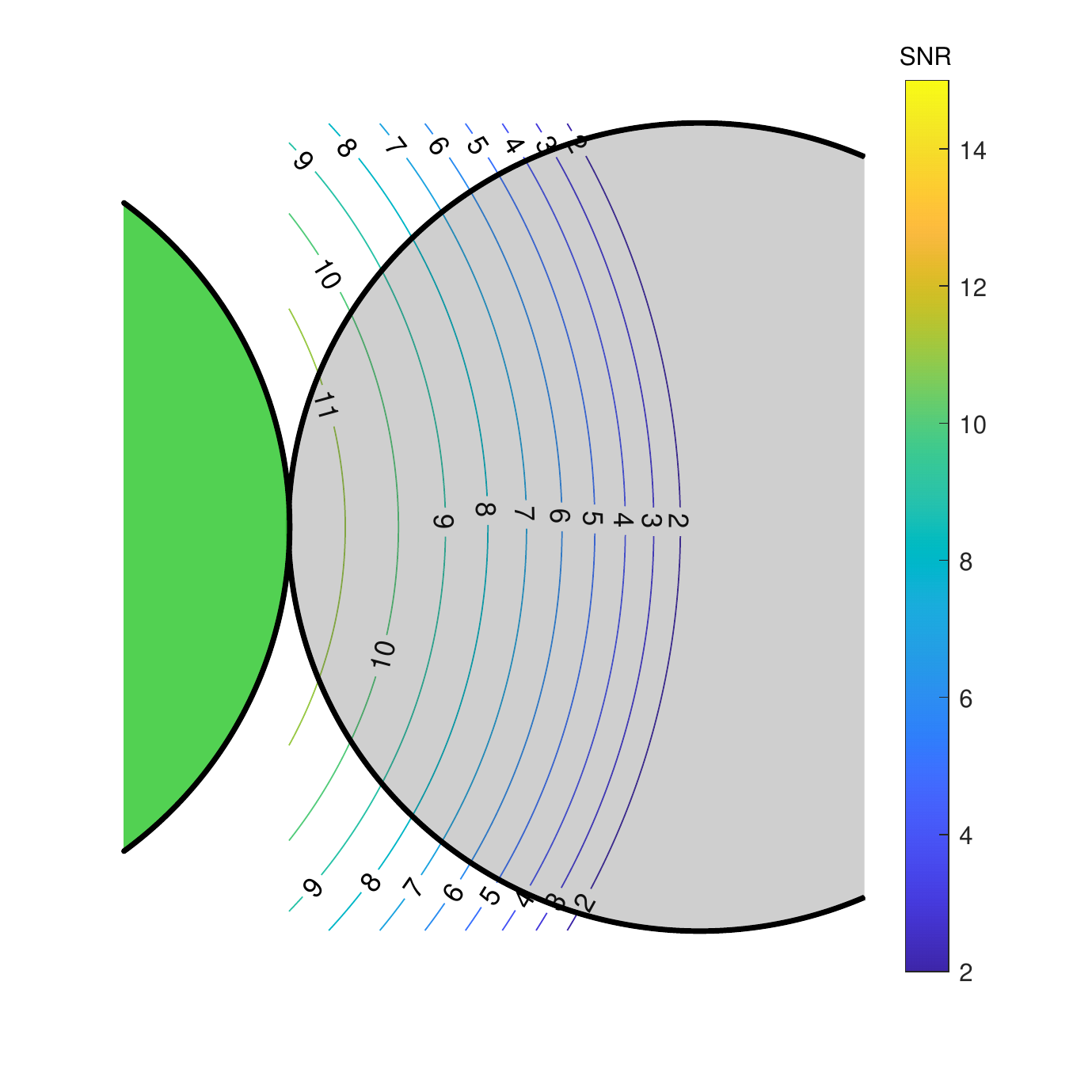}   \\ (a)  \\	\includegraphics[width=0.3\textwidth]{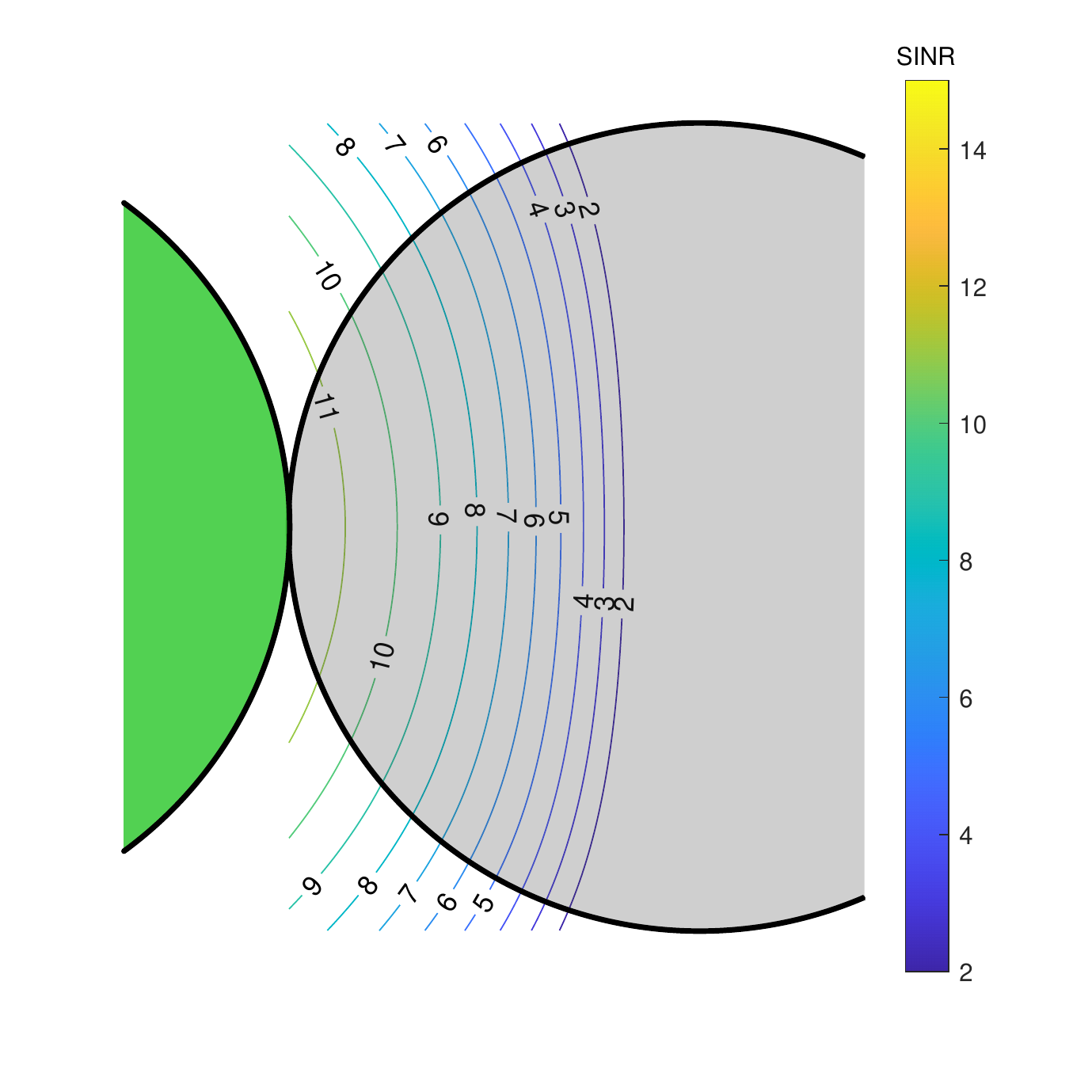}  \\  	 (b) 
		\end{tabular}
		\caption{ Contour map of the signal quality when  users are served by a non-dominant beam (green beam). The grey area displays the beam footprint at 3 dB: (a) SNR, (b) SINR.}
		\label{fig:snr_sinr_nodom}
	\end{figure}

\begin{figure}[!htb]
		\centering
			\includegraphics[width=0.3\textwidth]{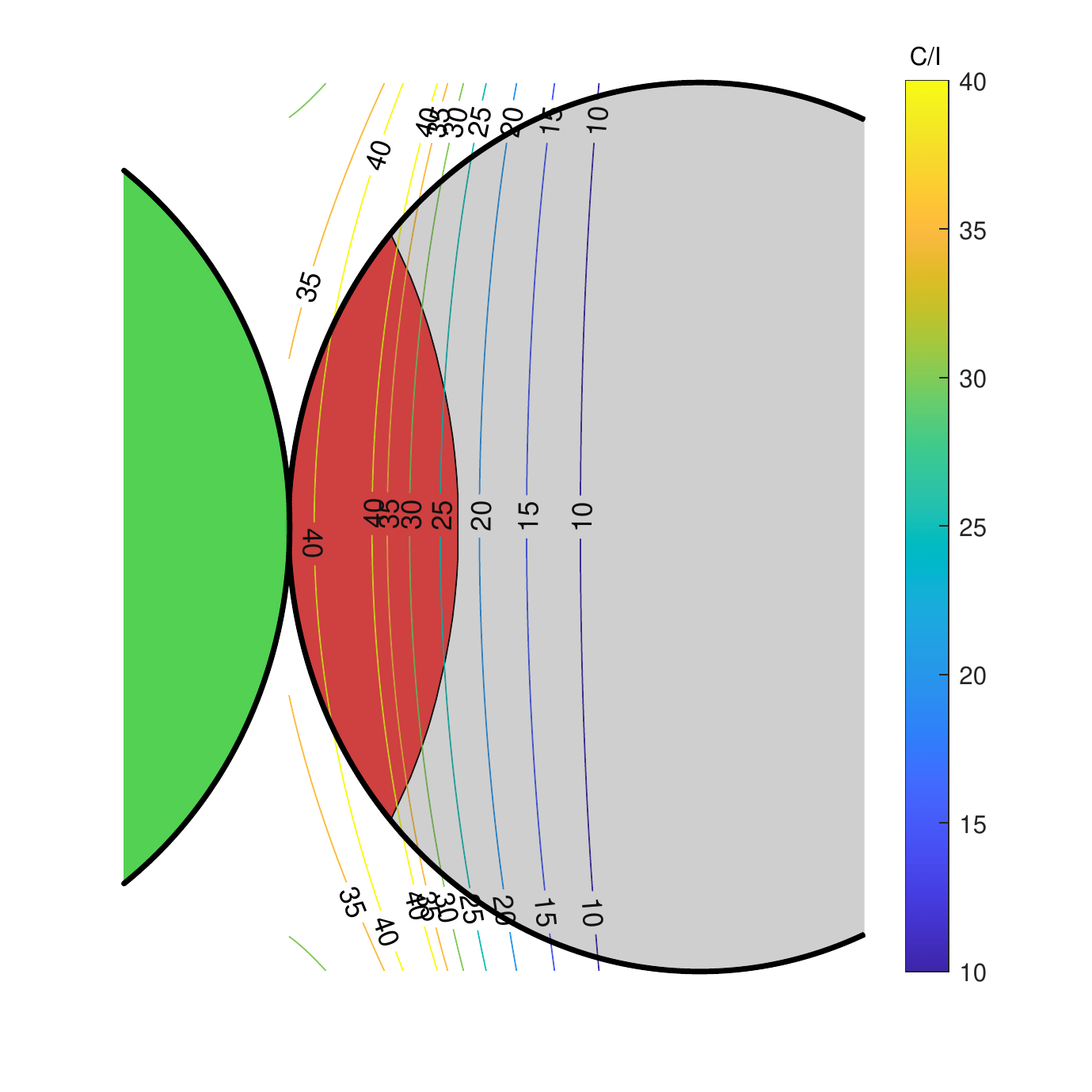}  	 
		\caption{ Contour map of the carrier to interference (C/I) ratio when  users are served by a non-dominant beam (green beam). The grey area displays the beam footprint at 3 dB. The inward red area  hosts those  users that  can be served by a non-dominant beam.}
		\label{fig:ci_nodom}
	\end{figure}

Following the described model in Section \ref{sec:system}, with a two-color pattern, the co-channel interference can be neglected, as justified next. For a uniform power and bandwidth allocation, the assumption holds when users are served by their dominant beam. This can been seen clearly in Fig. \ref{fig:snr_sinr_dom}, where  the signal-to-noise ratio (SNR) and signal-to-interference-and-noise ratio (SINR) are displayed for different user locations within a given cell. Note that the uniform allocation of frequency carriers, with a given carrier used every two beams, is the worst case  in terms of co-channel interference, since the uneven distribution of bandwidth across  beams can place interfering carriers in more distant beams. Quite the opposite, the peripheral areas of the cells can suffer from  higher  interference if power can be flexibly allocated, and skewed power beam distributions are applied as a result. In consequence, those solutions based on a flexible allocation of power can yield slightly optimistic results. 
Furthermore, the co-channel interference can limit the allocation of resources from neighbour beams, as displayed in Fig. \ref{fig:snr_sinr_nodom}. In order to keep the practical assumption on the interference, an SINR threshold is enforced and the resource pulling from neighbour beams is constrained to operate in the inward blue area  displayed in Fig. \ref{fig:ci_nodom}; this ensures a minimum carrier to interference (C/I) of $23$ dB with no significant impact on the SINR. For reference, the SNR at the center of the beam is approximately $15$ dB with the satellite parameters in Table \ref{tab:SystemPar}, which goes down to $8.7$ dB when users are served by a non-dominant beam \footnote{Current DVB-S2X standard can support this SINR range for the proposed flexible beam-user mapping.}. As additional information, the resource pulling area amounts to around $12\%$ of the overall beam area.

\subsection{ Traffic Model }

For reference purposes we will consider the satellite capacity, interpreted as the total average throughput that the satellite can deliver when resources (power and bandwidth) are uniformly allocated across beams. 
For the case under study, the 
capacity of the system is $T = 6.8$ Gbps for the six considered beams. This value is obtained by concentrating all the demand at each cell in a single user with SNR = 13.5 dB, which is the effective SNR for the system parameters in Table \ref{tab:SystemPar} and users uniformly distributed within the beam footprint. 

\begin{figure}[!htb]
		\centering
			\includegraphics[width=0.45\textwidth]{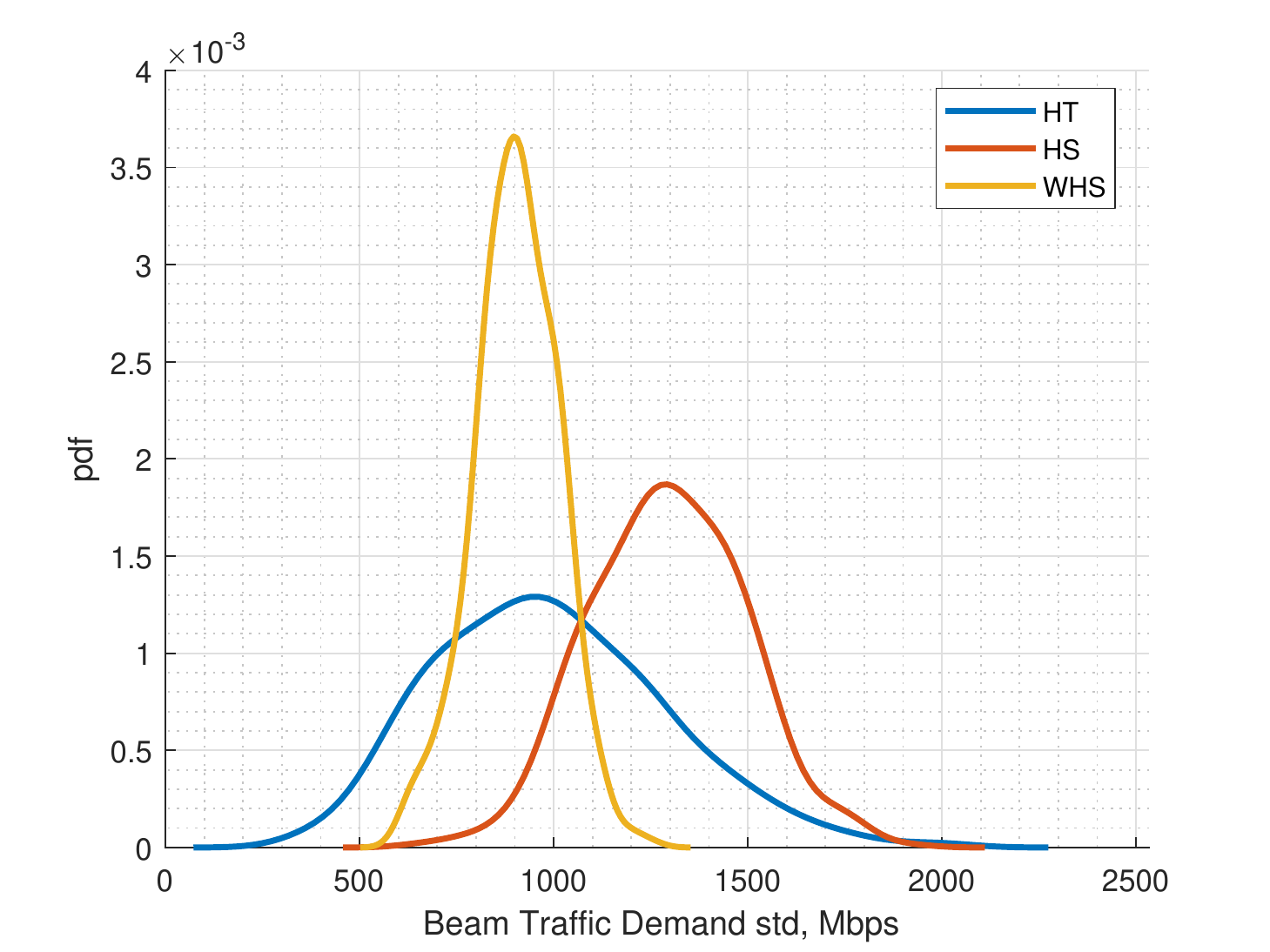}  	 
		\caption{Empirical probability distribution function of the standard deviation of the cells traffic demand, 500 realizations.}
		\label{fig:pdf_std}
	\end{figure}

We assume that every user demands a rate of  $25$ Mbps; in an attempt to  use up  the overall capacity $T$, we try to accommodate 272 users across the different beams. 
 We employ the  Dirichlet distribution  $Dir(K,\alpha)$, with $\bm \alpha=[\alpha_1,\dots, \alpha_K]$, to model the different traffic profiles
 while ensuring that the overall capacity $T$ is requested; the $\alpha_i$ parameters feature the possible statistically different traffic demands across cells. In particular, we portray  three different scenarios in this work: 
\begin{enumerate}[a)]
    \item \textit{Homogeneous traffic (HT)}:  A homogeneous exploration of the different traffic distributions is made with equal probability for every possible distribution of users among cells. This setting is modelled with $\alpha_i=1,\; i=1,\dots,K$.
    \item \textit{Hot-Spot (HS)}: A high amount of traffic is requested by users in one cell, with adjacent colder cells, in what is known as  the  HS scenario \cite{Nader2017,2018ICSSC,Tar18}. This non-uniform setting is modelled with $\bm \alpha=[5 \; 5 \; 30 \; 5 \; 5 \; 5]$, where the cell with index 3 is selected as the HS.
    \item \textit{Wide Hot-Spot (WHS)}: Another asymmetric scenario is explored, with two highly congested cells surrounded by colder cells. In this case, we place the high traffic demand in the footprints of beams  sharing the same HPA\footnote{Results for those schemes with adjustable power can differ slightly depending on the particular location of the pair of congested cells.}. Following the notation involving the HPAs in  Section \ref{sec:system}, this setting is modeled with $\bm \alpha=[10 \; 10 \; 40\; 40 \; 10 \; 10]$ by placing the congestion in the cells with indexes 3 and 4..

\end{enumerate}
The number of users per cell is obtained as follows. 
  First, $K$ random  numbers  are  taken  from  the Dirichlet distribution  $Dir(K,\alpha)$.  Then, they are multiplied by 272 and rounded accordingly. Next, users are uniformly distributed across the geographical locations of  each cell. 
  The traffic demand per cell, i.e.,  the number of users multiplied by $25$ Mbps, will be characterized by its standard deviation, ranging from 0 (uniform traffic demand) till $\sqrt{K-1}\frac{T}{K}$ (all traffic requested from a single  beam). 
For illustration purposes, the empirical probability distribution function of the standard deviation of the cells traffic demand  are presented in Fig.  \ref{fig:pdf_std}  for the  explored scenarios.  

\subsection{Numerical Results}

\begin{table*}[!h]
	\centering 
		\begin{tabular}{|c|c|c|c|c|c|}
			\hline
			\rowcolor[HTML]{C0C0C0} 
			\textbf{Technique} & \textbf{Label} & \textbf{ NQU} & \textbf{NU} & \textbf{  \makecell{	Offered \\ Rates \\ {[Gbps]}}  }  & \textbf{ \makecell{Minimum Rate \\ {[Mbps]} }  }  \\ \hline
			\multirow{2}{*}{ Beam Power Allocation} & \multirow{2}{*}{POW} & \multirow{2}{*}{0.162  }  & \multirow{2}{*}{ 0.337 }  & \multirow{2}{*}{ 4.511   }  & \multirow{2}{*}{ 10.41 }   \\  
				&  & &    &  & 	\\ \hline 
			\multirow{2}{*}{ Beam Bandwidth Allocation} & \multirow{2}{*}{BW}& \multirow{2}{*}{0.134  }  & \multirow{2}{*}{ 0.269 }  & \multirow{2}{*}{4.979}  & \multirow{2}{*}{6.59 }   \\  
				& &  &    &  & 	\\ \hline 
			\multirow{2}{*}{ \makecell{	Beam Power\&Bandwidth Allocation \\ (Benchmark) }} & \multirow{2}{*}{BW-POW} &  \multirow{2}{*}{0.088 }  & \multirow{2}{*}{0.218} & \multirow{2}{*}{ 5.323  }  & \multirow{2}{*}{9.41  }  \\  
			&   & &   &  & 	\\ \hline 
			\multirow{2}{*}{ \makecell{ Beam-user Mapping \\with fixed resources per beam  }  }& \multirow{2}{*}{MAP} &  \multirow{2}{*}{0.113  }  &  \multirow{2}{*}{0.248} &  \multirow{2}{*}{5.116 }   & \multirow{2}{*}{11.72 }   \\  
				&   & &   &  & 	\\ \hline 
			\multirow{2}{*}{ \makecell{ Beam-user Mapping \\  with flexible Bandwidth per Beam}  } & \multirow{2}{*}{BW-MAP}&   \multirow{2}{*}{0.084 }   & \multirow{2}{*}{ 0.219   } &  \multirow{2}{*}{5.312 }  & \multirow{2}{*}{12.63 }     \\
			&   &  &   &  & 	\\ \hline 
		\end{tabular}
	\caption{Average performance of the techniques in the HT scenario. The capacity of the satellite is $T=6.8$ Gbps, which is the aggregated rate requested by the users.} \label{tab:tech_res}
\end{table*}

As seen above, the quadratic unmet capacity in \eqref{eq:unmet2} is the baseline metric for optimization purposes. However, for a better understanding of the demand supply and comparison purposes, the following normalized parameters will be used: 
\begin{enumerate}
    \item \textit{Normalized quadratic unmet capacity: } 	\begin{align}
\text{NQU} &=   \frac{  \displaystyle \sum_{n=1}^N  (R^{req}(n) - R^{off}(n))^2    }{ N \left(W^{\sim} {\log_2(1+\snr_{eff})}\right)^2}
	\end{align}
with $\snr_{eff}$ an effective SNR corresponding to a super-user that embodies the user channels across all locations,  such that
     \begin{equation}\label{eq:snr_eff}
	\log_2(1+{\snr}_{eff}) =\mathbb{E}_h \left[ \log_2\left(1+ \frac{ P_{uni}^{\sim}  |h|^2} {N_o W^{\sim}} \right) \right]
	\end{equation}	
	where  $h$ is a random channel value, and the expectation is taken across all locations within the beam. Under this normalization, values close to 0 correspond to a better resource flexibility to provide the requested traffic, whereas higher values indicate the opposite.

	\item \textit{Normalized unmet capacity:}

		 \begin{equation}
\text{NU} =    \displaystyle \frac{T- \sum_{n=1}^N  R^{off}(n)  }{T}
	\end{equation}
	with $T$ the satellite capacity. This is a common metric found in the literature \cite{Wang2014,cocco2018,paris2019,Garau2020}. Note that for our simulations we are assuming that the aggregated rate of all users is equal to $T$, i.e., 
	\begin{equation}
	    T = \sum_{n=1}^N  R^{req}(n)
	\end{equation}
	since we are interested in studying the impact of the different load across beams.  
\end{enumerate}
In addition, the  minimum rate for each  simulated scenario will be also detailed to asses the fairness of the corresponding rate allocation scheme. 

As initial assessment of the different degrees of flexibility in the resource assignment, we address the HT scenario.  
The average performance of the different system metrics is collected in Table \ref{tab:tech_res}, where the normalized parameters NQU and NU are listed  along with the averaged offered and minimum rates.

  \begin{figure}[!htb]
		\centering
			\includegraphics[width=0.45\textwidth]{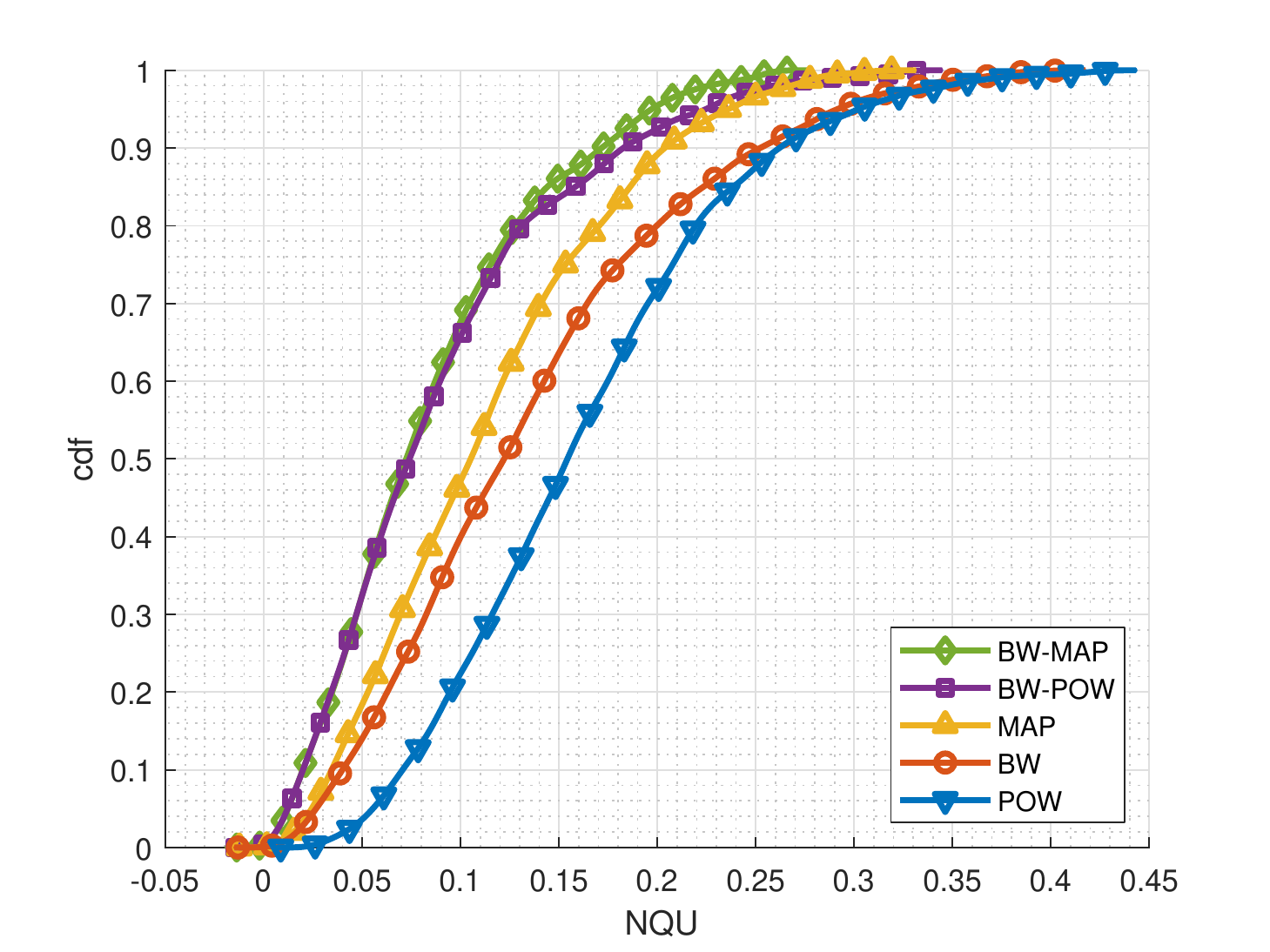} 	 
		\caption{ Cumulative distribution function of the normalized quadratic unmet capacity in the HT scenario.}
		\label{fig:cdf_NQU}
	\end{figure}
	
A first relevant observation is that the adjustability when mapping beams and users, or allocating power,  enhances similarly the performance of bandwidth flexibility. 
 This can be seen clearly in Fig. \ref{fig:cdf_NQU}, that displays the Cumulative Distribution Function (CDF) of the normalized quadratic unmet capacity. Note, however, that the implications of both schemes are different; on the one side, an  ideal  scenario  is  considered  for  the  flexible  power allocation and more practical constraints should be enforced for the power amplification \cite{cocco2018}: a change in the operation point of a power amplifier modifies its power efficiency and the intermodulation interference caused by  the non-linearity. On the other side, the complexity of the associated optimization is significantly different, with a genetic algorithm used to solve the joint power and bandwidth allocation problem. In this regard, some attempts can be found in the literature to conceive a convex optimization problem as an approximation, see, e.g. \cite{Wang2014,lagunas2020flexible_jour}.
 
 \begin{figure}[!htb]
		\centering
			\includegraphics[width=0.45\textwidth]{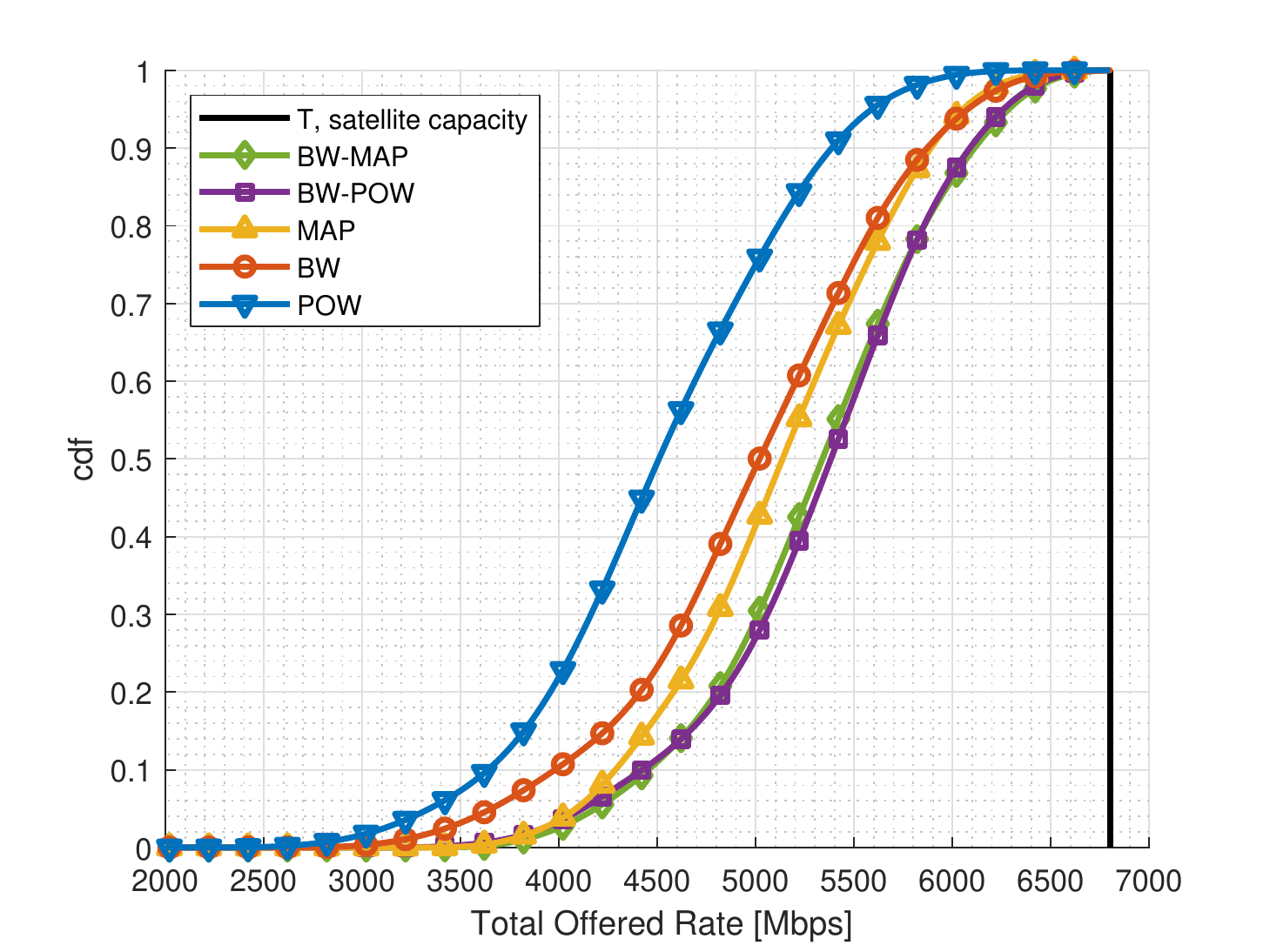} 	 
		\caption{ Cumulative distribution function of the total offered rate in the HT scenario.}
		\label{fig:cdf_Off}
	\end{figure}

 Remarkably,  we can also see from Fig. \ref{fig:cdf_NQU} that under a fixed allocation of the payload resources to the different beams,  the flexible beam-user mapping by itself  can provide a competitive performance against the conventional rigid mapping with adjustable beam bandwidth. In addition, let us note that the lack of competitiveness of power flexible solutions agrees with previous results in the literature \cite{cocco2018,paris2019}.

In terms of the offered rates, previous conclusions hold, with similar  performance of  both  beam-user mapping and flexible power allocation when pairing together with a flexible allocation of  bandwidth, enhancing the performance of the latter around  6\%, as concluded from Table \ref{tab:tech_res}.   
A detailed view is presented in Fig. \ref{fig:cdf_Off}, where the  CDF of the total offered rates for the different techniques is displayed.  
 
\begin{figure}[!htb]
		\centering
			\includegraphics[width=0.45\textwidth]{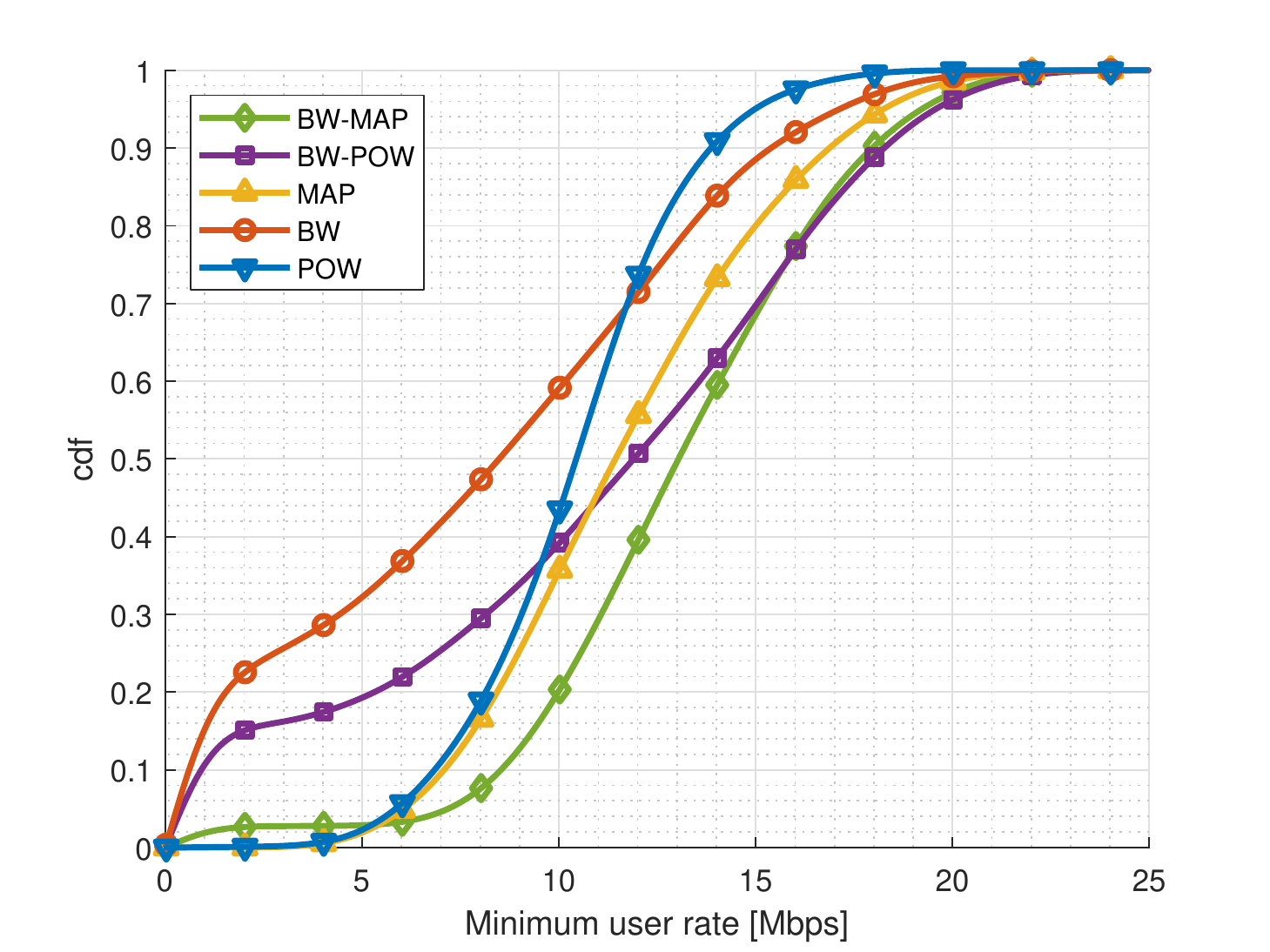}	 
		\caption{ Cumulative distribution function of the minimum rates in the HT scenario.}
		\label{fig:cdf_Min}
	\end{figure}

 Fairness in the provision of rates is also relevant; in particular, the average minimum rate for all users is displayed in Table \ref{tab:tech_res}, with
 the corresponding CDF in Fig. \ref{fig:cdf_Min}. We can see that the  flexible bandwidth combines better, from a fairness perspective, with the smart beam-user mapping rather than the  adjustable power allocation,  achieving higher probabilities for intermediate minimum rates.
 
		\begin{figure}[!htb]
		\centering
			\includegraphics[width=0.45\textwidth]{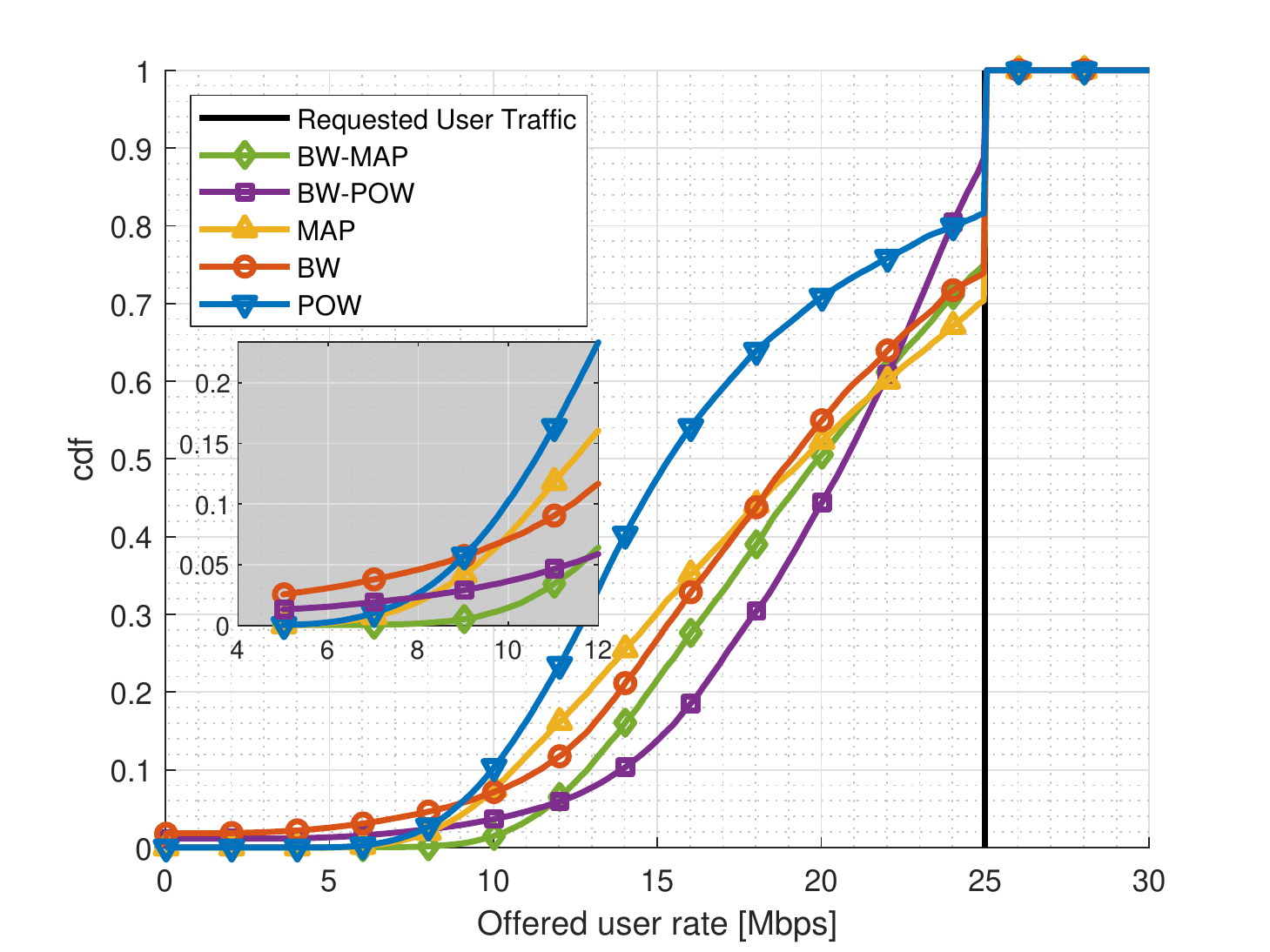}	 
		\caption{ Cumulative distribution function of the offered user rates in the HT scenario.}
		\label{fig:cdf_UserRates}
	\end{figure}

 In addition,  the CDF of all user rates across the explored traffic demand distribution is presented in Fig. \ref{fig:cdf_UserRates}.  Despite the similar performance in terms of the average quadratic unmet capacity in Table \ref{tab:tech_res},  both flexible power allocation  and smart mapping of the users pursue the provision of the user demand differently when paired together with the flexible bandwidth. By inspecting  Fig. \ref{fig:cdf_UserRates}, the former exchanges both lower and higher ends of the user rates  for more users with intermediate rates, as compared with the latter. The flexible beam user-mapping raises the floor with a 9\% of improvement on the average minimum rate in Table \ref{tab:tech_res}, whereas granting full rate to  more users, in particular 
 13\% higher by looking at the cross of both  curves with the requested user traffic abscissa in Fig.  \ref{fig:cdf_UserRates}. Finally, note  in the same figure that the flexible beam-user mapping enhances notably the performance of flexible bandwidth when combined, as  early hinted in \cite{Ram21}.

\begin{table*}[!h]
	\centering 
		\begin{tabular}{|c|c|c|c|c|c|c|}
			\hline
			\rowcolor[HTML]{C0C0C0} 
			\textbf{ Case } & \textbf{Technique} & \textbf{Label}  &\textbf{ NQU} & \textbf{NU} & \textbf{  \makecell{	Offered \\ Rates \\ {[Gbps]}}  }  & \textbf{ \makecell{Minimum Rate \\ {[Mbps]} }  }  \\ \hline
			 \multirow{3}{*}{HS} & Rigid mapping & BW  &  0.253     &  0.388    &  4.164     & 0.88        \\  \cline{2-7}
			 & Power Flexibility & BW-POW &   0.189    &  0.321    &  4.624     &  2.47       \\  \cline{2-7}
				  & Beam-user Mapping & BW-MAP &   0.126   &  0.284     &  4.874      &  10.72   \\  \hline
			\multirow{3}{*}{WHS} & Rigid Mapping & BW & 0.172    &  0.336    &  4.515     & 9.5     \\  \cline{2-7}
			 & Power Flexibility  & BW-POW &  0.110     &   0.278    & 4.913       & 12.70      \\ \cline{2-7}
				  & Beam-user Mapping & BW-MAP &  0.112    &    0.256   &  5.059      &  12.13      \\  \hline
	
		\end{tabular}
	\caption{Average performance of the explored techniques in the Wide Hot-Spot (WHS) and Hot-Spot (HS) scenarios.} \label{tab:perf_sel}
\end{table*}

\begin{figure}[!htb]
		\centering
		\subfloat[User location for a given realization. The circles represents different user locations, and are filled with a color denoting the beam which serves the corresponding user 
		  in the case of flexible beam-user mapping combined with  flexible bandwidth.\label{fig:Ex_2}]{%
         \includegraphics[width=0.45\textwidth]{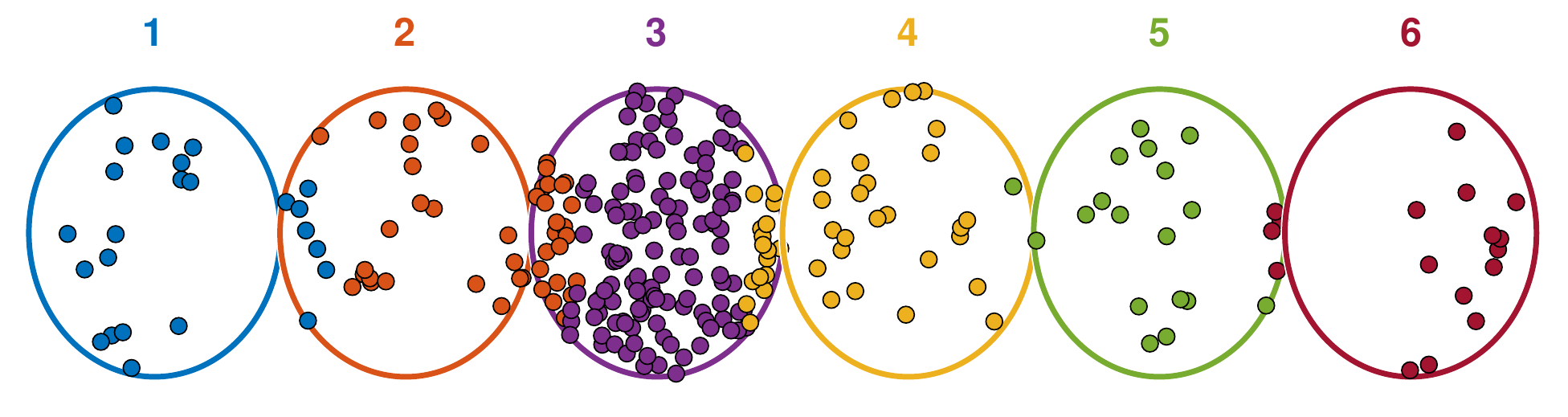}} 
    \hfill
  \subfloat[ Average requested and offered traffic per beam.\label{fig:HS_Traffic}]{%
         \includegraphics[width=0.45\textwidth]{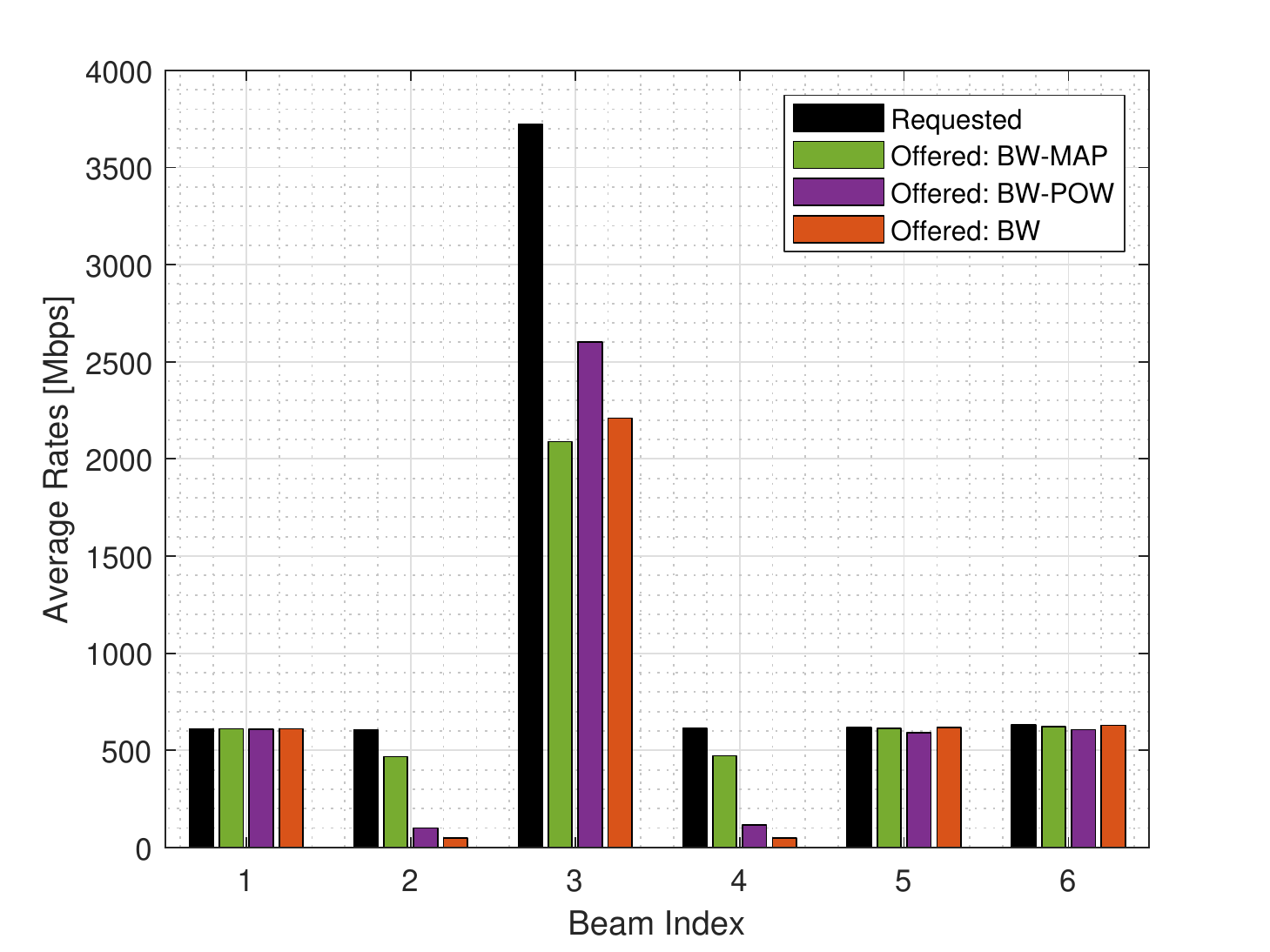} }
		\caption{Hot-Spot scenario.}
		\label{fig:HS}
	\end{figure}

\begin{figure}[!h]
		\centering
			\includegraphics[width=0.45\textwidth]{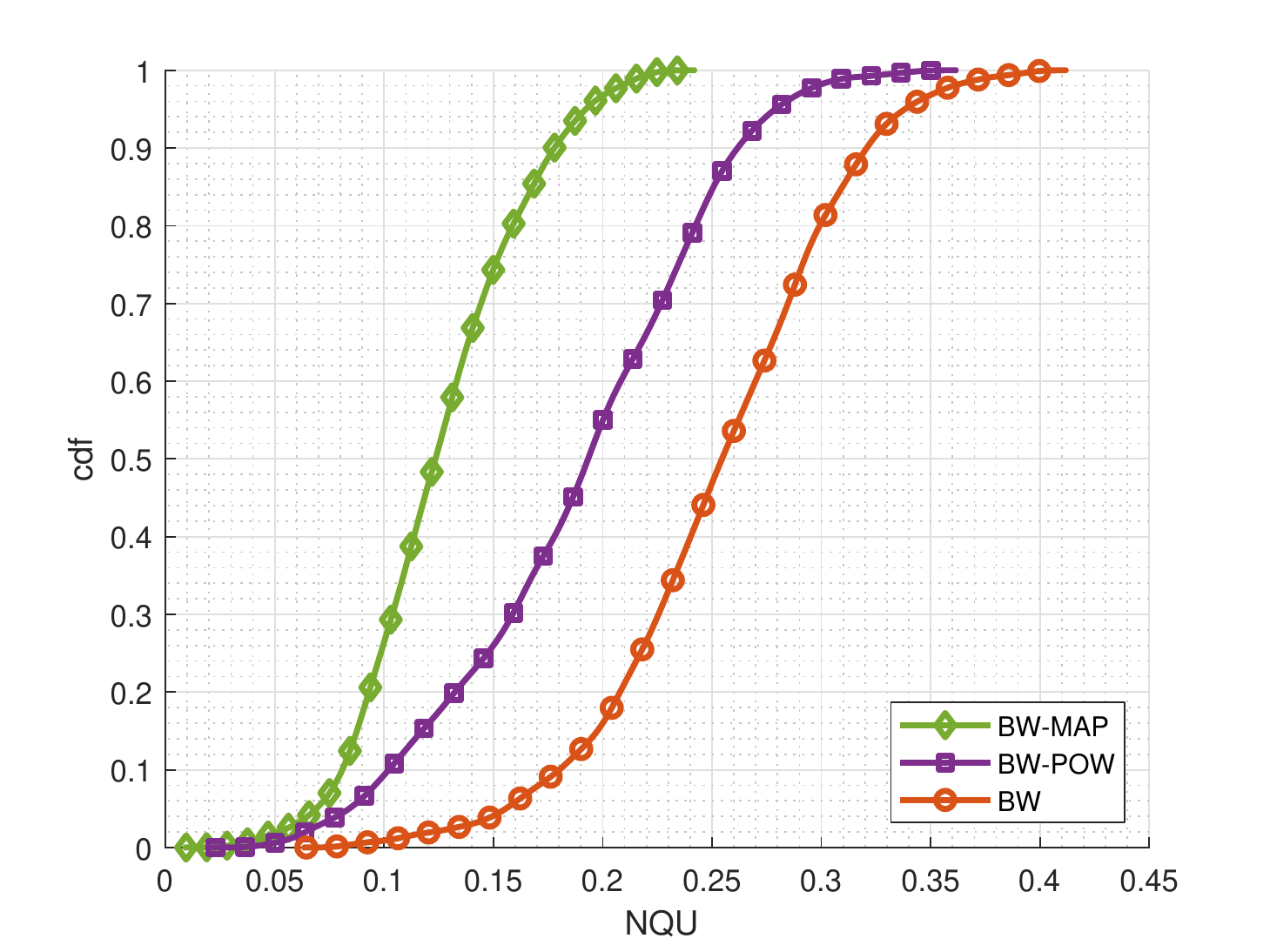}	 
		\caption{ Cumulative distribution function of the normalized quadratic unmet capacity in the Hot-Spot (HS) scenario. }
		\label{fig:cdf_NQU_HS}
	\end{figure}

\begin{figure}[!h]
		\centering
			\includegraphics[width=0.45\textwidth]{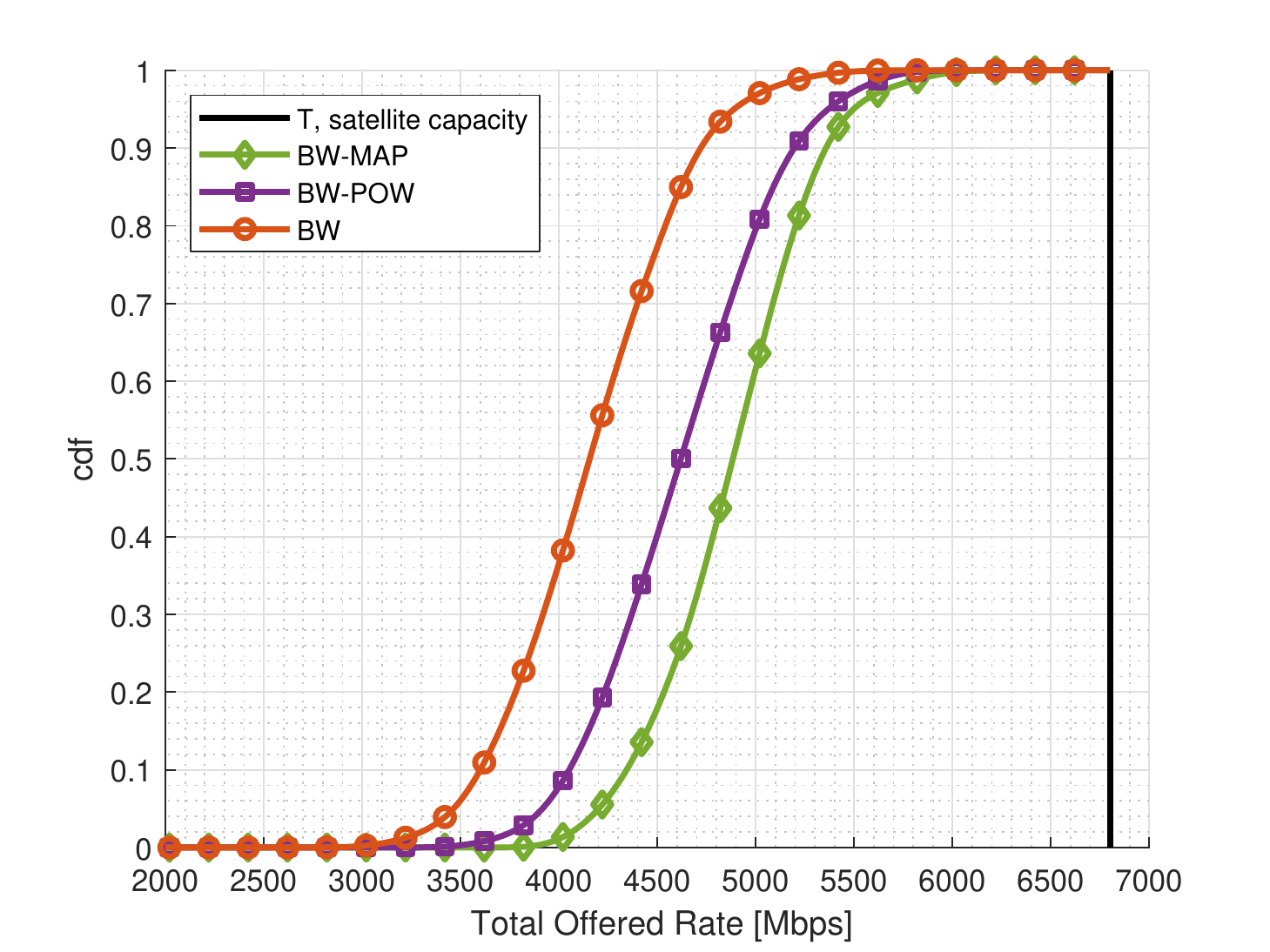}
		\caption{ Cumulative distribution function of the total offered rate in the  Hot-Spot (HS) scenario.}
		\label{fig:cdf_OfferRates_HS}
	\end{figure}

\begin{figure}[!h]
		\centering
			\includegraphics[width=0.45\textwidth]{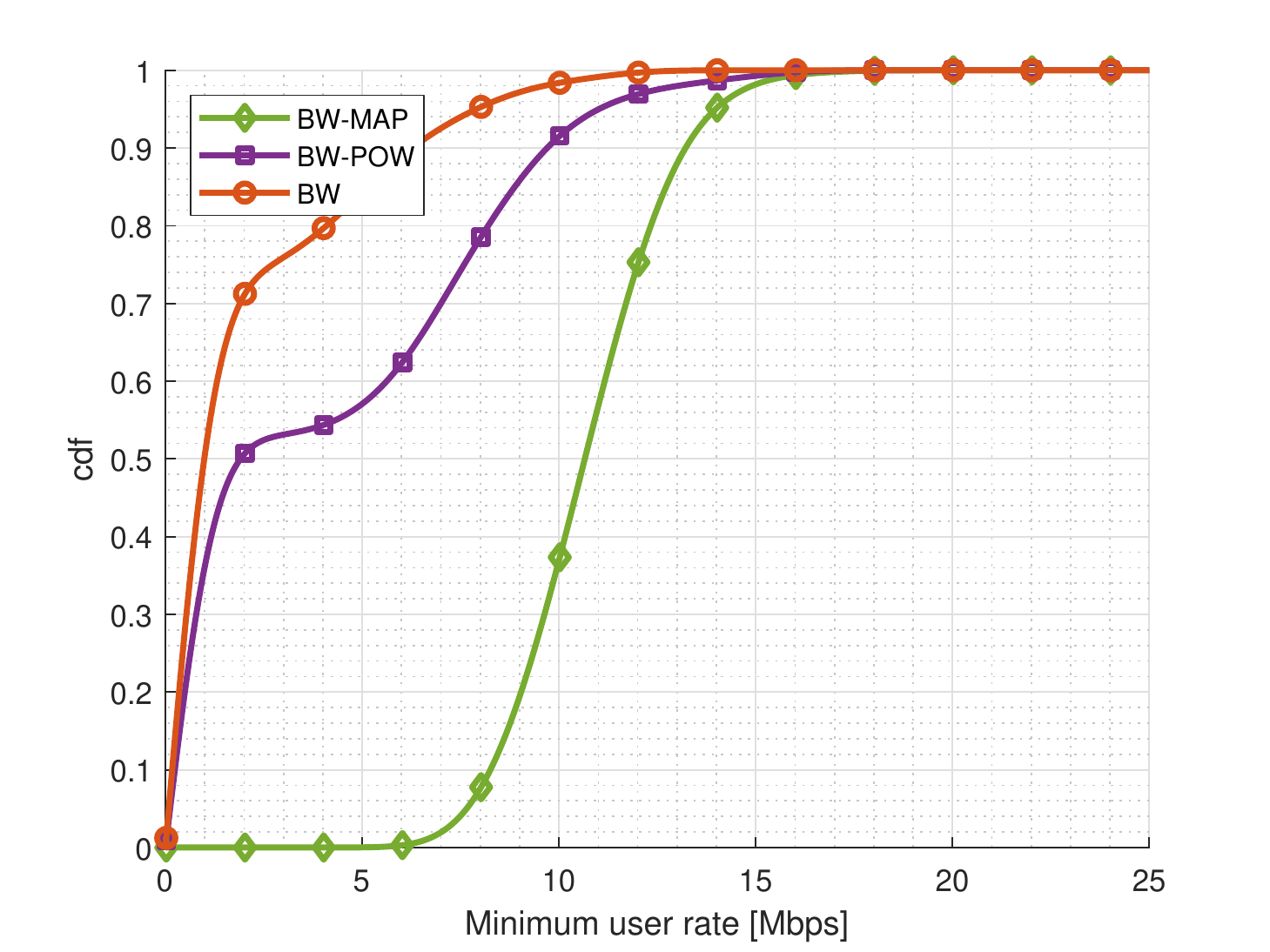}	 
		\caption{ Cumulative distribution function of the minimum rates in the Hot-Spot (HS) scenario.}
		\label{fig:cdf_MinRates_HS}
	\end{figure}

\begin{figure}[!h]
		\centering
			\includegraphics[width=0.45\textwidth]{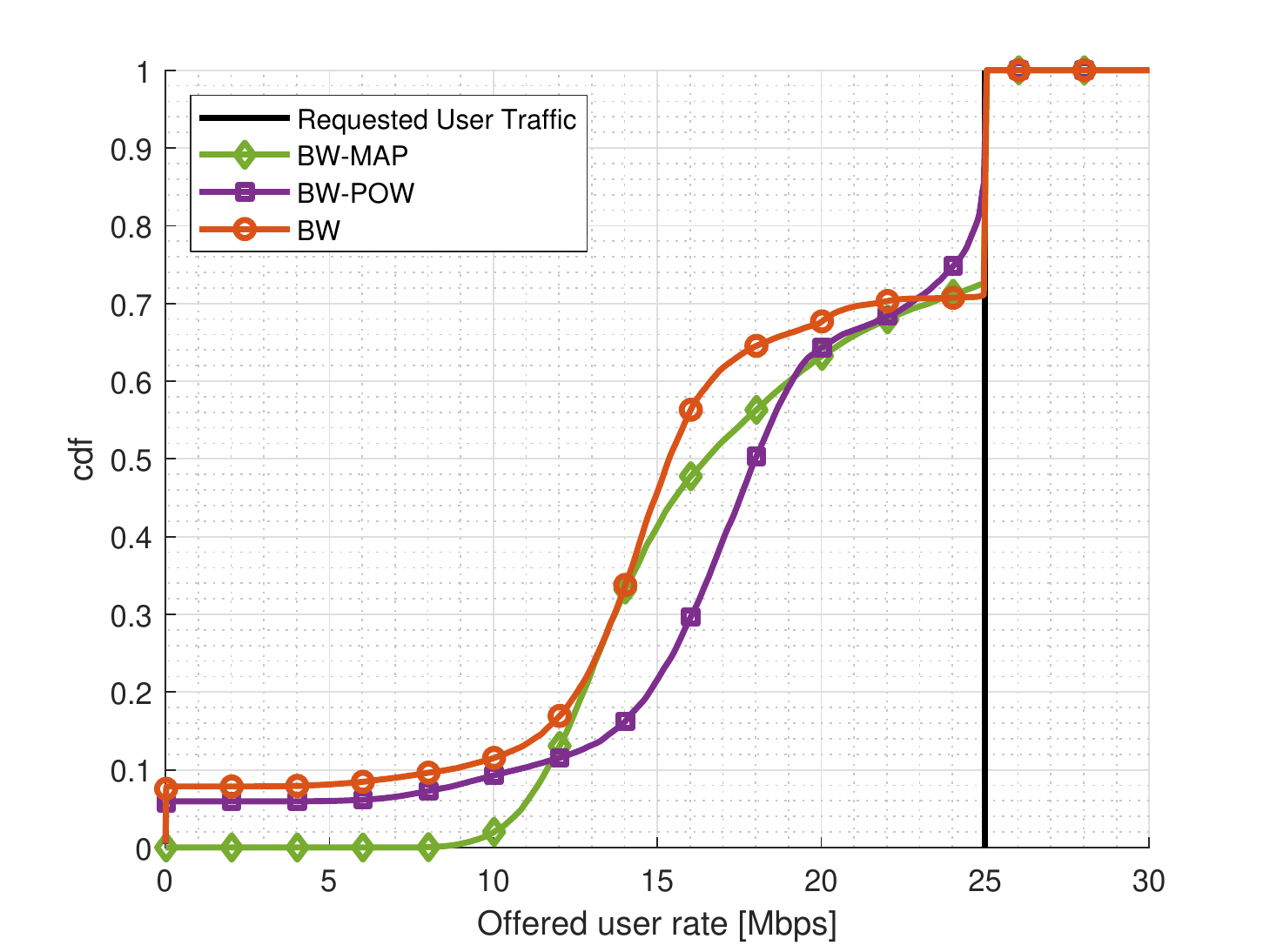}
		\caption{ Cumulative distribution function of the offered user rates in the Hot-Spot (HS) scenario.}
		\label{fig:cdf_UserRates_HS}
	\end{figure}

So far, we have addressed diverse traffic profiles in the HT scenario without favouring any particular placement of the traffic among cells. However, the flexible beam-user mapping can particularly excel under strongly skewed traffic distributions. To this end, the techniques under comparison are also tested in two different asymmetric scenarios. First, we address an especially skewed case, the  Hot-Spot (HS) scenario, with one congested beam, surrounded by colder beams;  the corresponding numerical results are collected in  Table \ref{tab:perf_sel}. For the sake of clarity, we discard those schemes with fixed bandwidth allocation, and label the techniques in terms of their additional flexibility on top of the flexible bandwidth allocation hereafter. For illustration purposes, an example of user distribution is sketched in Fig. \ref{fig:Ex_2}, with the average requested and offered traffic per beam  presented in Fig.  \ref{fig:HS_Traffic}. Note that, in this HS case, both  rigid and flexible mapping solutions devote most of the resources to the congested beam,  depriving almost completely the two colder adjacent beams of any offered throughput. By just allowing an alternative link, the flexible beam-user mapping can trade off some offered traffic in the congested beam for a  better match of the traffic demand in the colder beams, with an improvement on the data provision around 5\% and 17\%, on average, over the adjustable power allocation and rigid mapping, respectively. Remarkably, this improvement is obtained in a cooperative way, as it can be observed from the example in Fig. \ref{fig:Ex_2}, and the smart beam-user mapping results in a joint effort to supply the traffic demand with multiple beams involved in  the resource pulling, some of them far away from  the concentration of high traffic demand. As a matter of fact, the flexible association between users and beams leads to better matching of the traffic demand with a reduction on the average quadratic unmet demand around 33\%  and 50\% against the power flexibility and rigid mapping,   with substantial improvements on the minimum rates in both cases.  The improvement with the flexible-beam user mapping is evident from the CDF curves of the normalized quadratic unmet capacity, total offered rates, minimum user rates and offered user rates in Fig. \ref{fig:cdf_NQU_HS}, \ref{fig:cdf_OfferRates_HS}, \ref{fig:cdf_MinRates_HS} and \ref{fig:cdf_UserRates_HS}, respectively. It is highly remarkable the improvement on the minimum rate for the HS scenario provided by the BW-MAP scheme, as per Table \ref{tab:perf_sel} and Fig. \ref{fig:cdf_MinRates_HS}. This comes from a  more fair allocation or resources to those beams adjacent to the congested one, which in turn receives a slightly lower rate as compared to other schemes (see Fig. \ref{fig:HS}).

\begin{figure}[!htb]
		\centering
		\subfloat[User location for a given realization. The circles represents different user locations, and are filled with a color denoting the beam which serves the corresponding user 
		  in the case of flexible beam-user mapping combined with  flexible bandwidth.\label{fig:Ex_1_A}]{%
         \includegraphics[width=0.45\textwidth]{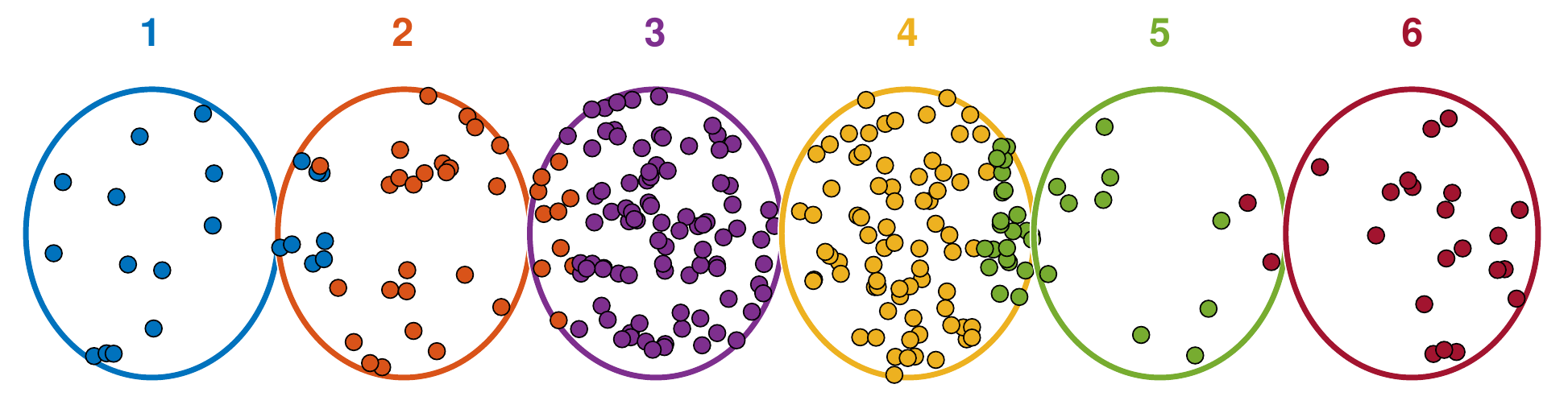}}
    \hfill
  \subfloat[ Average requested and offered traffic per beam\label{fig:WHS_A_Traffic}]{%
          \includegraphics[width=0.45\textwidth]{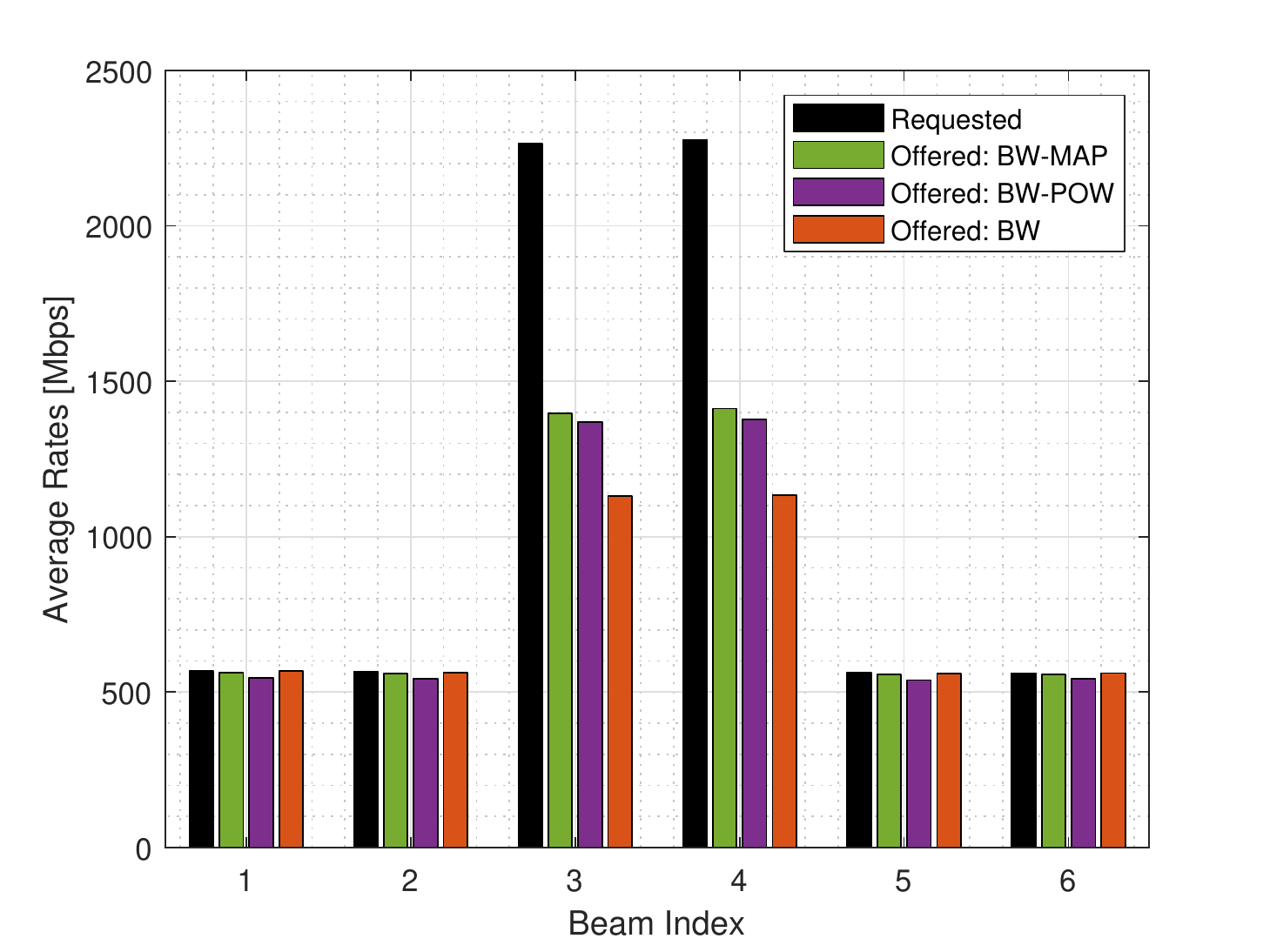} }
		\caption{Wide Hot-Spot scenario.}
		\label{fig:WHS}
	\end{figure}

\begin{figure}[!h]
		\centering
			\includegraphics[width=0.45\textwidth]{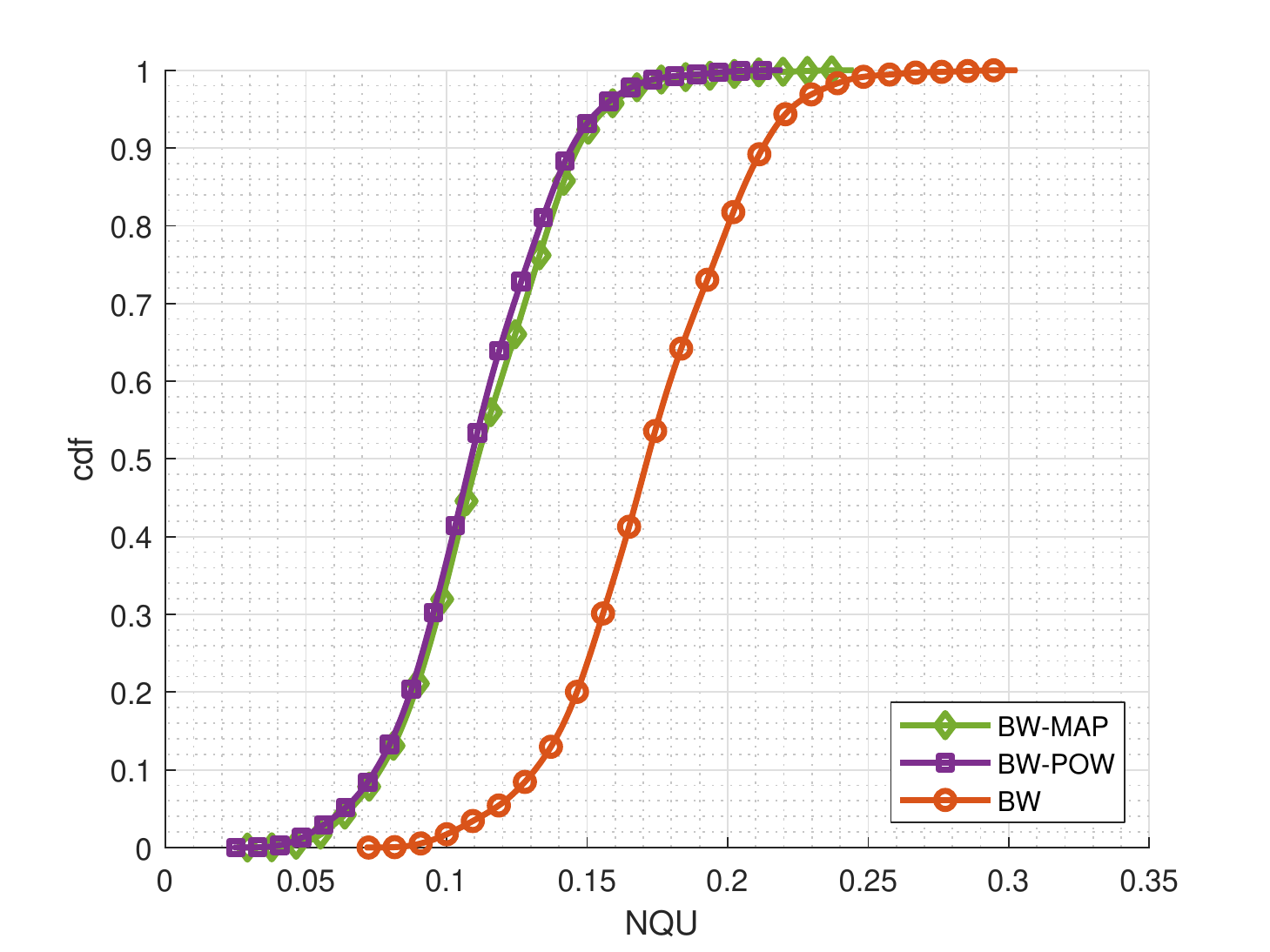}	 
		\caption{ Cumulative distribution function of the normalized quadratic unmet capacity in the  Wide Hot-Spot (WHS) scenarios.}
		\label{fig:cdf_NQU_WHS}
	\end{figure}

\begin{figure}[!h]
		\centering
			\includegraphics[width=0.45\textwidth]{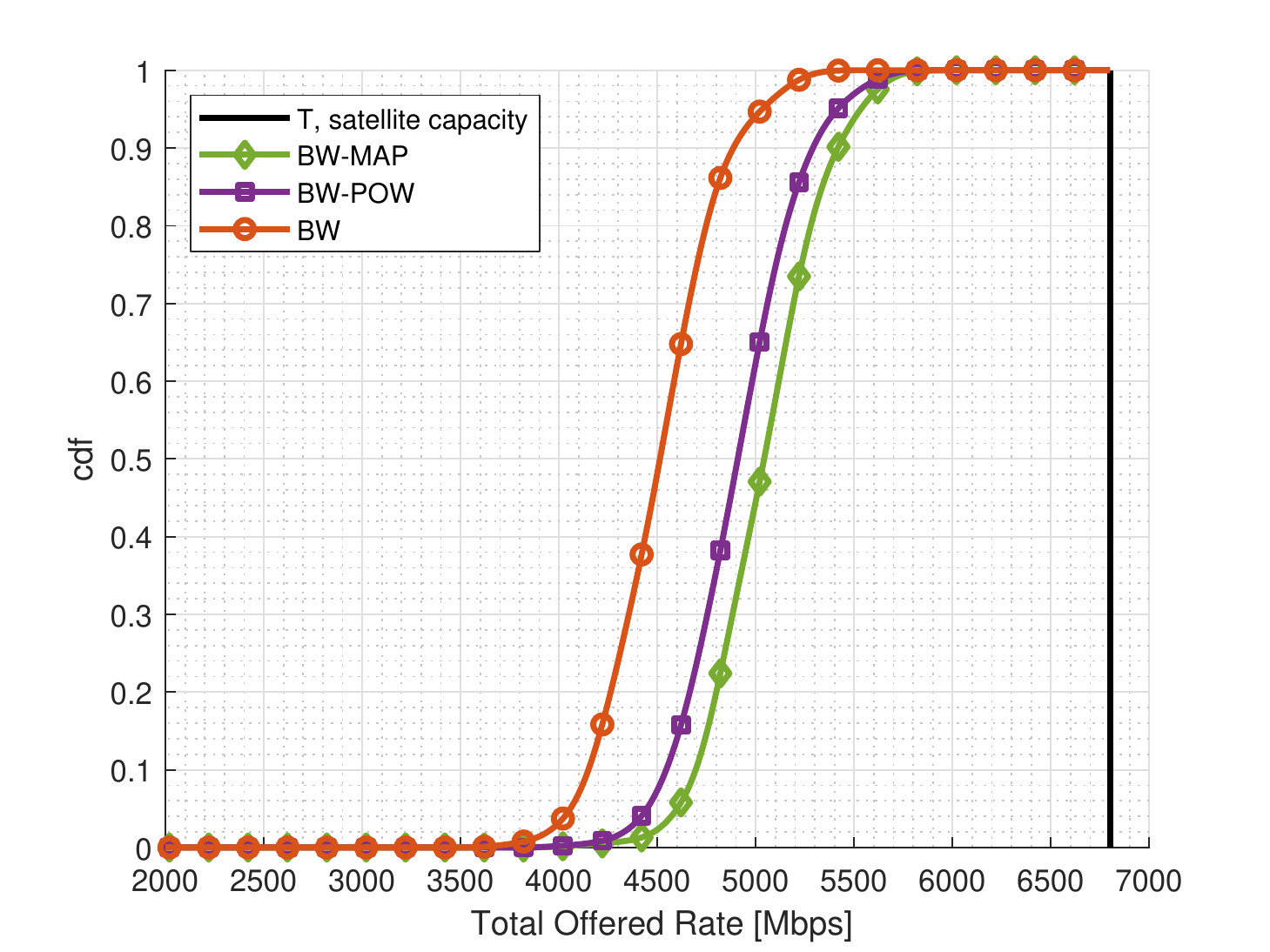}
		\caption{ Cumulative distribution function of the total offered rate in the Wide Hot-Spot (WHS) scenarios.}
		\label{fig:cdf_OfferRates_WHS}
	\end{figure}

\begin{figure}[!h]
		\centering
			\includegraphics[width=0.45\textwidth]{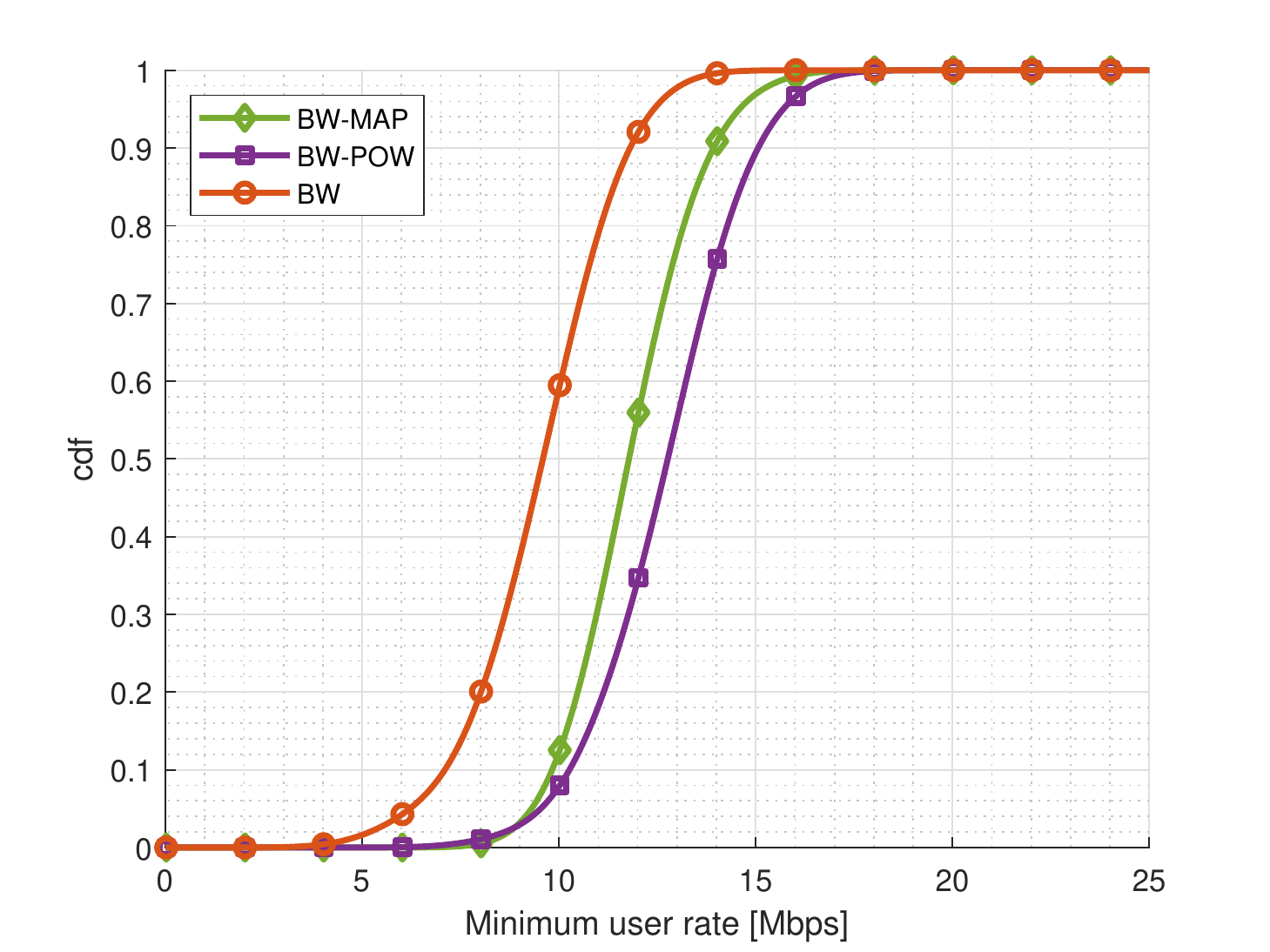}	 
		\caption{ Cumulative distribution function of the minimum rates in the Wide Hot-Spot (WHS) scenarios.}
		\label{fig:cdf_MinRates_WHS}
	\end{figure}

\begin{figure}[!h]
		\centering
			\includegraphics[width=0.45\textwidth]{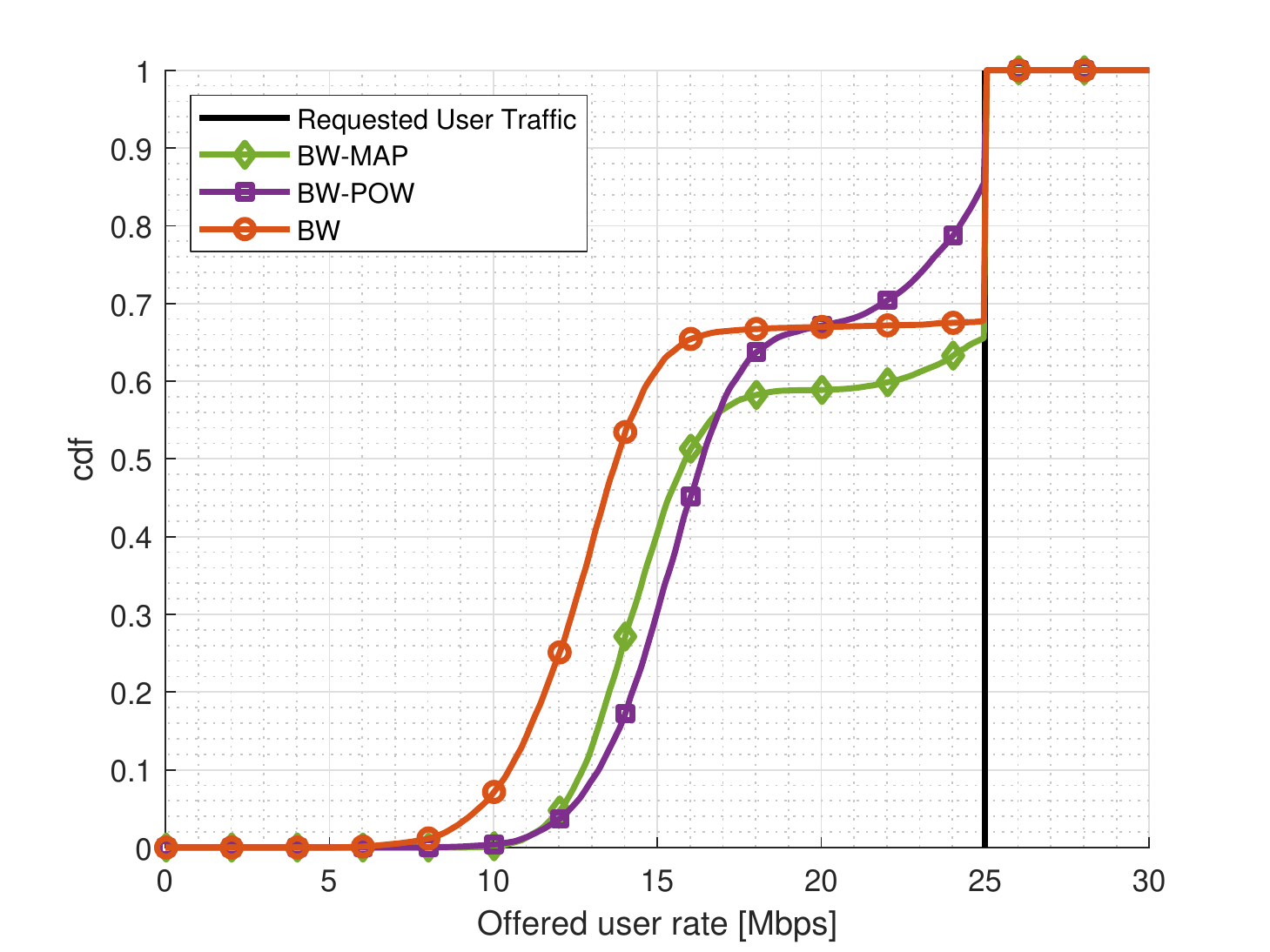}
		\caption{ Cumulative distribution function of the offered user rates in the Wide Hot-Spot (WHS) scenarios.}
		\label{fig:cdf_UserRates_WHS}
	\end{figure}

 Finally, we consider the WHS scenario, with two strongly congested beams  surrounded by colder beams, as sketched in   Fig. \ref{fig:Ex_1_A} for a given  realization. The average  requested and offered rates per beam are depicted in Fig. \ref{fig:WHS_A_Traffic}, with performance metrics detailed in  Table \ref{tab:perf_sel}. 
  From this table, the flexible beam-user mapping provides a reduction of around 35\%  in terms of quadratic unmet demand against the rigid beam-user mapping in the WHS scenario;  the improvement is clear from the CDF of the  normalized quadratic unmet demand in Fig. \ref{fig:cdf_NQU_WHS}. This results in both better demand satisfaction and fairness in the system.  The smart mapping of the users increases the demand satisfaction, close to 12\%, and also raises the floor of the user rates with increments of the minimum user rate of approximately 28\% against the rigid mapping. This better performance can be seen in Figs. \ref{fig:cdf_OfferRates_WHS}, \ref{fig:cdf_MinRates_WHS} and \ref{fig:cdf_UserRates_WHS}, which display the CDF of the total offered rates, minimum user rates and offered user rates, respectively. Note that the overlapping bandwidth constraint limits the number of carriers available for the users in congested cells in caes of rigid mapping. Interestingly, the flexible beam-user mapping overcomes these limitations through a smart association of  users and beams, and performing a cooperative provision of the user rates. On the other hand,
  the adjustable power allocation presents a close performance to the flexible beam-user mapping  in terms of quadratic unmet demand as reflected in Fig. \ref{fig:cdf_NQU_WHS}.
  However, we  should be keep in mind that the performance with  adjustable power allocation is somewhat optimistic, so that   enforcing more realistic power constraints would yield an additional edge for the  flexible beam-user mapping solution.

 As a final remark, let us note that the  performance evaluation in this section was obtained for a fixed satellite antenna radiation diagram, and it is left 
 for future studies the joint optimization of both user assignment and beamforming. As an illustration, hot-spot like scenarios can benefit from a non-rigid  steering of the satellite beams, as pointed out in \cite{2021ICSSC}. Advanced adaptive beamforming capabilities, if available, can be employed to modulate  the beams load and contribute to a smoother service, even in the absence of flexible RRM. Additionally, flexible  beam-user mapping can provide load transferring capabilities without requiring the use of  active antennas. Furthermore, both load managing approaches are not exclusive and can be jointly optimized. Thus, the flexible beam-user mapping could simplify the  beamforming requirements for transferring the traffic load.

 \section{Conclusions} \label{sec:Con}

 This work has explored the improvement margin from a flexible pairing between users and beams, with particular emphasis on flexible payloads, for which radio resource management is instrumental to serve spatially non-uniform traffic demands in satellite multibeam footprints. Under the proposed user-centric approach, a two-step optimization process has revealed as a practical strategy, with a first convex optimization problem allocating carriers, power and/or users to the different beams. A second optimization step, operating at beam level, assigns carriers to users. The Dirichlet distribution has been chosen to feature different random traffic demands, due to its capability to shape the  traffic profile across beams. 
   From the results, it can be concluded that a non-rigid association between users and beams, operating jointly with a flexible allocation of bandwidth, can compete with a joint power-bandwidth allocation, even with ideal conditions and lenient constraints for the power allocation. This comes without noticeable additional complexity at the different subsystems, and with more simple optimization schemes. Remarkably, for those cases for which traffic is severely asymmetric across beams, in hot-spot like cases, the use of a flexible beam-user mapping can report significant benefits when serving the requested traffic levels.

\section*{Acknowledgment}
  Funded by the Agencia Estatal de Investigaci\'on (Spain) and the European Regional Development Fund (ERDF) through the project RODIN 
 (PID2019-105717RB-C21). Also funded by Xunta de Galicia (Secretaria Xeral de Universidades) under a predoctoral scholarship (cofunded by the European Social Fund). 
 The views of the authors of this paper do not necessarily reflect  the views of the European Space Agency.

\appendix 

\section{Convexity proof of the beam power optimization}
\label{sec:appA}

We can write $g(\bm x)$ in \eqref{eq:opt-power2} as the sum of separable functions, so that the optimization problem for the vector $\bm x = (x_1,\ldots,x_K)$ of fractions of total power per beam reads as 
  \begin{equation}\label{eq:opt-power3}
     \begin{array}{ll}
       \mbox{max} & g(\bm x) = \frac{1}{2} \left(\frac{W^{total}}{2}\right)^2  \\ &  \hphantom{ g(\bm x) = \frac{1}{2} } \cdot  \sum_{b=1}^K \frac{1}{|\mathcal N(b)|} \left( f_b(x_b) - g_b(x_b) \right)  \\ \\
      	 \mbox{subject to}    &    \sum_{b \in A(j)} x_b  \leq \frac{P^{max}}{P^{total}}  ,\; j=1,\ldots,K/2      \\   
 	 \hphantom{ \mbox{subject to}}  &   \sum_{b=1}^K x_b  \leq 1 
\end{array}
\end{equation}
with
\begin{eqnarray}
  f_b(x) & \triangleq &  2 S^{req}(b)\log_2\left(1+ x \cdot {\snr}_b\right) \\
  g_b(x) & \triangleq & \log_2^2\left(1+x \cdot {\snr}_b \right)
\end{eqnarray}
and the requested spectral efficiency $S^{req}(b) = R^{req}(b)/(W^{total}/2)$.

It can be proved that $f_b(x)-g_b(x)$ is a quasiconcave function for positive $x$; unfortunately, the sum of quasiconcave functions is not guaranteed to be quasiconcave, which would help to come up with an efficient optimization algorithm. In fact, it can be checked that the sum in \eqref{eq:opt-power3} is not necessarily quasiconcave in all cases. Nevertheless, concavity of the function $f_b(x)-g_b(x)$ can be analyzed in detailed; its second derivative is given by 
\begin{multline}
  \frac{\partial^2}{\partial x^2}(f_b(x) - g_b(x)) = \\ -\text{cst}
  \frac{S^{req}(b) - \log_2(1+x\cdot {\snr}_b) +  \log_2 e}{(1+x\cdot {\snr}_b)^2}
\end{multline}
with $\text{cst}$ a positive constant. Since the offered traffic must be lower than the requested traffic, i.e., $\log_2(1+x\cdot {\snr}_b) \leq S^{req}(b)$, we have that the second derivative is negative for the range of admissible values of $x$ and, in consequence,  problem \eqref{eq:opt-power3} and, equivalently, \eqref{eq:opt-power2}, is  convex. \\

\section{Genetic Algorithm implementation}
\label{sec:appB}
 
In a  genetic algorithm, individuals evolve from previous solutions with the aim to optimize a given objective function. This evolution process is composed of four main operations:
\begin{itemize}
    \item \textbf{Mutation}:  The attributes of a individual are randomly modified. When generating a new generation, there is  a probability $p_{mut}$  of a individual to be affected by this mutation. 
    \item \textbf{Selection of the fittest}: In each iteration, an operation is made to select the  parents to generate the new individuals in the next generation. 
      \item \textbf{Crossover:} The characteristics of two parents are stochastically combined to generate new individuals. This operations is applied with a probability $p_{cross}$ by selecting two random parents. 
      \item \textbf{Elite members}: A determined number of the best solutions are set as new individuals for the next generation and remain unchanged.
\end{itemize}

\begin{figure}[!htb]
		\centering
			\includegraphics[width=0.48\textwidth]{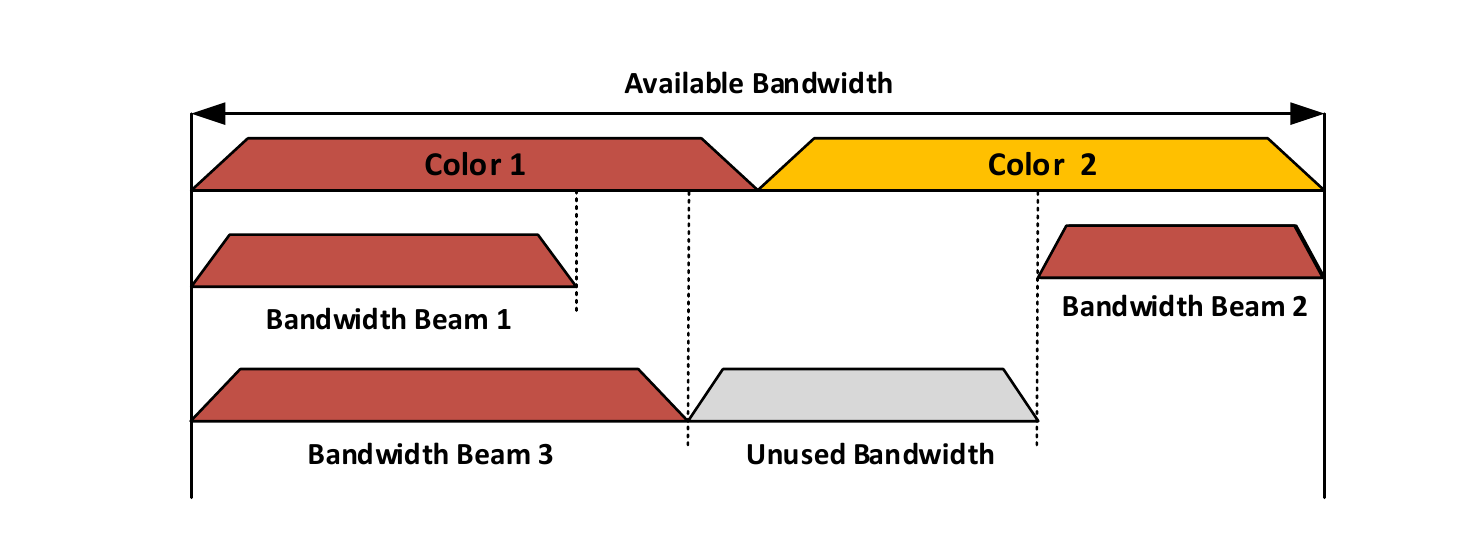} 
		\caption{ Example for which the available bandwidth is not fully allocated to beams 1, 2 and 3. The color of beam 2 is different from the color of beams 1 and 2 \cite{paris2019}.}
		\label{fig:B_un}
	\end{figure}

For the genetic algorithm design, we follow a similar implementation to \cite{paris2019}, with the bandwidth and power per beam as individuals of the algorithm. For the benchmark solution, we define $\bm C = [M_1,\ldots,M_K] \in  \mathbb{Z}^{K}$ as the vector that collects the number of carriers per beam $M_b$. Then, we set  the pair of vectors $\bm P$ and $\bm C$ as the inputs of the algorithm. The evolution process is driven by the  combinations of  $\bm P$ and $\bm C$, which have to be ranked. We employ the  geometric mean of the user channels under the footprint of the beam $b$ to obtain a reference channel $h_{b}^{ref}$. Then, the requested and offered traffic in the beam $b$ are computed as
  \begin{align}
  	R_{beam}^{req}(b)  &=   \sum_{n \in {\mathcal N}(b)}  R^{req}(n)  \\
    R^{off}_{beam}(b)  &= W_b \log_2\left(1+  \frac{P_b|h_{b}^{ref}|^2}{N_0 W_b}  \right) ,\quad W_b= M_b W^\sim
  \end{align}
  and the quadratic unmet function is evaluated as 
    \begin{equation}
   \mathcal U  =  \sum_{b=1}^K \left(	R_{beam}^{req}(b) -R^{off}_{beam}(b)  \right)^2.
  \end{equation}
For further details of the algorithm implementations,  Table \ref{tab:Gen_Set} collects the settings employed by the genetic algorithm; particular crossover and mutation functions are employed for  integer inputs \cite{GA_Int}. In addition, an in-between process is considered,  as the randomly generated individuals do not always satisfy  the overall power constraint and beam bandwidth overlapping constraints. To resolve this, the same approach  as in \cite{paris2019} is made to repair the incorrect solutions to satisfy the constraints:
\begin{itemize}
    \item \textbf{ Power constraints}:  To satisfy the sum power constraint, the vector $\bm P$ is scaled down by a factor $k$. If the sum power of a vector $\bm P$ is given by $P_{sum}$ and it is higher than $P^{total}$, then the power is scaled down by $k= \frac{P^{total}}{P_{sum}}$. In addition, the power is also modified to satisfy the HPA power constraints: $  \sum_{b \in A(j)} P_b  \leq P^{max}  ,\; j=1,\ldots,K/2$.
    \item \textbf{Bandwidth overlapping constraint}:  To avoid the overlapping frequencies between two neighbour beams, the bandwidth constraint C3
    in \eqref{eq:Gen_Opt} has to be satisfied. After selecting randomly an increasing or descending order of the beams, the bandwidth constraint is enforced in an iterative process that  goes through all possible beam indexes $b$. If  $W_b + W_{d} > W^{total}$ during the process, with $d=\{b-1,b+1 \}$ depending on the selected order, then we set $W_b=W^{total}-W_{d}$ to satisfy the overlapping bandwidth constraint with $W_b= M_b  W^\sim$. 
    However, this process can produce an inefficient bandwidth allocation with unused portions of bandwidth (see Fig. \ref{fig:B_un}).  For a given triplet of beams, the unused bandwidth $W_{un}$ is expressed as 
    \begin{equation}
        W_{un} = W^{total} - W_c -\max( W_l, W_r)
    \end{equation}
    where $c,l,r$ are the central, left and right beam indices of the triplet. Different central beams can be chosen, so that the beams with higher traffic demand are considered first, and the corresponding  unused bandwidth $W_{un}$ is assigned to them, till no residual unused bandwidth is left.

\end{itemize}

\begin{table}[!htb]
	\centering
	\begin{tabular}{|c|c|}
		\hline
		\rowcolor[HTML]{C0C0C0} 
		Parameter & Value \\  \hline 
	   	Selection operator &  Tournament 	\\    
	    Crossover operator &   Laplace crossover \\   
	   	Mutation operator &  Power mutation \\   \hline  
	   Max. number of generations  & 5000  	\\   
	   Population size  &  4000 	\\   
	   Tournament size  &  5 	\\      
	   Elite members &  20	\\ \hline 
	   Crossover prob. & 0.8 \\  
	   Mutation prob. & 0.1\\ \hline 
	\end{tabular}
	\caption{Genetic algorithm settings\cite{GA_Int}.}\label{tab:Gen_Set}
\end{table}



\bibliography{ref}

\begin{thebibliography}{10}
\providecommand{\url}[1]{#1}
\csname url@samestyle\endcsname
\providecommand{\newblock}{\relax}
\providecommand{\bibinfo}[2]{#2}
\providecommand{\BIBentrySTDinterwordspacing}{\spaceskip=0pt\relax}
\providecommand{\BIBentryALTinterwordstretchfactor}{4}
\providecommand{\BIBentryALTinterwordspacing}{\spaceskip=\fontdimen2\font plus
\BIBentryALTinterwordstretchfactor\fontdimen3\font minus
  \fontdimen4\font\relax}
\providecommand{\BIBforeignlanguage}[2]{{%
\expandafter\ifx\csname l@#1\endcsname\relax
\typeout{** WARNING: IEEEtran.bst: No hyphenation pattern has been}%
\typeout{** loaded for the language `#1'. Using the pattern for}%
\typeout{** the default language instead.}%
\else
\language=\csname l@#1\endcsname
\fi
#2}}
\providecommand{\BIBdecl}{\relax}
\BIBdecl

\bibitem{Ang2021}
P.~Angeletti and J.~L. Cubillos, ``Traffic balancing multibeam antennas for
  communication satellites,'' \emph{IEEE Transactions on Antennas and
  Propagation}, pp. 1--1, 2021.

\bibitem{Park12}
U.~Park, H.~W. Kim, D.~S. Oh, and B.~J. Ku, ``{Flexible Bandwidth Allocation
  Scheme Based on Traffic Demands and Channel Conditions for Multi-Beam
  Satellite Systems},'' in \emph{2012 IEEE Vehicular Technology Conference (VTC
  Fall)}, 2012, pp. 1--5.

\bibitem{choi2005}
J.~P. Choi and V.~W. Chan, ``{Optimum power and beam allocation based on
  traffic demands and channel conditions over satellite downlinks},''
  \emph{{IEEE} Trans. Wireless Commun.}, vol.~4, no.~6, pp. 2983--2993, 2005.

\bibitem{Qi15}
T.~Qi and Y.~Wang, ``{Energy-efficient power allocation over multibeam
  satellite downlinks with imperfect CSI},'' in \emph{2015 International
  Conference on Wireless Communications Signal Processing (WCSP)}, 2015, pp.
  1--5.

\bibitem{Efen20}
C.~N. Efrem and A.~D. Panagopoulos, ``{Dynamic Energy-Efficient Power
  Allocation in Multibeam Satellite Systems},'' \emph{IEEE Wireless Commun.
  Lett.}, vol.~9, no.~2, pp. 228--231, 2020.

\bibitem{Lei10}
J.~Lei and M.~A. V\'azquez-Castro, ``{Joint Power and Carrier Allocation for
  the Multibeam Satellite Downlink with Individual SINR Constraints},'' in
  \emph{2010 IEEE International Conference on Communications}, 2010, pp. 1--5.

\bibitem{Wang2014}
A.~L. Heng~Wang and X.~Pan, ``{Optimization of Joint Power and Bandwidth
  Allocation in Multi-Spot-Beam Satellite Communication Systems},'' \emph{Math.
  Probl. Eng.}, vol. 214, 2014.

\bibitem{cocco2018}
G.~Cocco, T.~De~Cola, M.~Angelone, Z.~Katona, and S.~Erl, ``{Radio Resource
  Management Optimization of Flexible Satellite Payloads for DVB-S2 Systems},''
  \emph{{IEEE} Trans. Broadcast.}, vol.~64, no.~2, pp. 266--280, 2018.

\bibitem{paris2019}
A.~Paris, I.~Del~Portillo, B.~Cameron, and E.~Crawley, ``{A Genetic Algorithm
  for Joint Power and Bandwidth Allocation in Multibeam Satellite Systems},''
  in \emph{2019 IEEE Aerospace Conference}.\hskip 1em plus 0.5em minus
  0.4em\relax IEEE, 2019, pp. 1--15.

\bibitem{lagunas2020flexible_jour}
T.~S. Abdu, S.~Kisseleff, E.~Lagunas, and S.~Chatzinotas, ``{Flexible Resource
  Optimization for GEO Multibeam Satellite Communication System},'' \emph{IEEE
  Trans. Wirel. Commun.}, pp. 1--1, 2021.

\bibitem{anzalchi2010}
J.~Anzalchi, A.~Couchman, P.~Gabellini, G.~Gallinaro, L.~D'agristina,
  N.~Alagha, and P.~Angeletti, ``{Beam hopping in multi-beam broadband
  satellite systems: System simulation and performance comparison with
  non-hopped systems},'' in \emph{2010 5th Advanced Satellite Multimedia
  Systems Conference and the 11th Signal Processing for Space Communications
  Workshop}.\hskip 1em plus 0.5em minus 0.4em\relax IEEE, 2010, pp. 248--255.

\bibitem{Lei11}
M.~A. Lei, Jiang V\'azquez-Castro, ``{Multibeam satellite frequency/time
  duality study and capacity optimization},'' \emph{J. Commun. Netw.}, vol.~13,
  no.~5, pp. 472--480, 2011.

\bibitem{Alberti10}
X.~Alberti, J.~M. Cebrian, A.~Del~Bianco, Z.~Katona, J.~Lei, M.~A.
  Vazquez-Castro, A.~Zanus, L.~Gilbert, and N.~Alagha, ``{System capacity
  optimization in time and frequency for multibeam multi-media satellite
  systems},'' in \emph{2010 5th Advanced Satellite Multimedia Systems
  Conference and the 11th Signal Processing for Space Communications Workshop},
  2010, pp. 226--233.

\bibitem{lei2020beam}
L.~Lei, E.~Lagunas, Y.~Yuan, M.~G. Kibria, S.~Chatzinotas, and B.~Ottersten,
  ``{Beam Illumination Pattern Design in Satellite Networks: Learning and
  Optimization for Efficient Beam Hopping},'' \emph{IEEE Access}, vol.~8, pp.
  136\,655--136\,667, 2020.

\bibitem{Xu2020}
X.~{Hu}, Y.~{Zhang}, X.~{Liao}, Z.~{Liu}, W.~{Wang}, and F.~M. {Ghannouchi},
  ``{Dynamic Beam Hopping Method Based on Multi-Objective Deep Reinforcement
  Learning for Next Generation Satellite Broadband Systems},'' \emph{{IEEE}
  Trans. Broadcast.}, vol.~66, no.~3, pp. 630--646, 2020.

\bibitem{Vaz18}
M.~A. V{\'a}zquez and A.~I. P{\'e}rez-Neira, ``{Spectral Clustering For
  Beam-Free Satellite Communications},'' in \emph{2018 IEEE Global Conference
  on Signal and Information Processing (GlobalSIP)}, 2018, pp. 1030--1034.

\bibitem{Vaz20}
------, ``{Multigraph Spectral Clustering for Joint Content Delivery and
  Scheduling in Beam-Free Satellite Communications},'' in \emph{ICASSP 2020 -
  2020 IEEE International Conference on Acoustics, Speech and Signal Processing
  (ICASSP)}, 2020, pp. 8802--8806.

\bibitem{ICASSP20}
T.~{Ramírez} and C.~{Mosquera}, ``{Resource Management in the Multibeam
  NOMA-based Satellite Downlink},'' in \emph{IEEE International Conference on
  Acoustics, Speech and Signal Processing (ICASSP)}, 2020, pp. 8812--8816.

\bibitem{Nader2017}
N.~Alagha, ``{'Adjacent Beams Resource Sharing to Serve Hot Spots'},'' in
  \emph{35th AIAA International Communications Satellite Systems Conference,
  International Communications Satellite Systems Conferences (ICSSC)}, Oct
  2017.

\bibitem{2018ICSSC}
T.~Ram{\'\i}rez, C.~Mosquera, M.~Caus, A.~Pastore, N.~Alagha, and N.~Noels,
  ``{Adjacent Beams Resource Sharing to Serve Hot Spots: A Rate-Splitting
  Approach},'' in \emph{36th International Communications Satellite Systems
  Conference (ICSSC 2018)}.\hskip 1em plus 0.5em minus 0.4em\relax IET, 2018,
  pp. 1--8.

\bibitem{Tar18}
G.~Taricco and A.~Ginesi, ``{Precoding for Flexible High Throughput Satellites:
  Hot-Spot Scenario},'' \emph{{IEEE} Trans. Broadcast.}, pp. 1--8, 2018.

\bibitem{Icolari2017}
V.~Icolari, S.~Cioni, P.-D. Arapoglou, A.~Ginesi, and A.~Vanelli-Coralli,
  ``{Flexible Precoding for Mobile Satellite System Hot Spots},'' in \emph{2017
  IEEE International Conference on Communications (ICC)}.\hskip 1em plus 0.5em
  minus 0.4em\relax IEEE, 2017, pp. 1--6.

\bibitem{Ram21}
T.~Ramirez, C.~Mosquera, and N.~Alagha, ``{Flexible Beam-User Mapping for
  Multibeam Satellites},'' in \emph{29th European Signal Processing Conference
  (EUSIPCO)}, 2021.

\bibitem{2021ICSSC}
T.~Ram{\'\i}rez, C.~Mosquera, and N.~Alagha, ``{Non-Orthogonal Transmission
  Under Flexible Illumination Patterns For Advanced Satellite Payloads},'' in
  \emph{38th International Communications Satellite Systems Conference (ICSSC
  2021)}.\hskip 1em plus 0.5em minus 0.4em\relax IET, 2021.

\bibitem{Garau2020}
J.~J.~G. Luis, N.~Pachler, M.~Guerster, I.~del Portillo, E.~Crawley, and
  B.~Cameron, ``{Artificial Intelligence Algorithms for Power Allocation in
  High Throughput Satellites: A Comparison},'' in \emph{2020 IEEE Aerospace
  Conference}.\hskip 1em plus 0.5em minus 0.4em\relax IEEE, 2020, pp. 1--15.

\bibitem{maral2020satellite}
G.~Maral, M.~Bousquet, and Z.~Sun, \emph{{Satellite communications systems:
  systems, techniques and technology}}.\hskip 1em plus 0.5em minus 0.4em\relax
  John Wiley \& Sons, 2020.

\bibitem{Bessel}
C.~{Caini}, G.~E. {Corazza}, G.~{Falciasecca}, M.~{Ruggieri}, and
  F.~{Vatalaro}, ``{A spectrum- and power-efficient EHF mobile satellite system
  to be integrated with terrestrial cellular systems},'' \emph{{IEEE} J. Sel.
  Areas Commun.}, vol.~10, no.~8, pp. 1315--1325, 1992.

\bibitem{aravanis2015}
A.~I. Aravanis, B.~S. MR, P.-D. Arapoglou, G.~Danoy, P.~G. Cottis, and
  B.~Ottersten, ``{Power Allocation in Multibeam Satellite Systems: A Two-Stage
  Multi-Objective Optimization},'' \emph{{IEEE} Trans. Wireless Commun.},
  vol.~14, no.~6, pp. 3171--3182, 2015.

\bibitem{lagunas2020CA}
M.~G. {Kibria}, E.~{Lagunas}, N.~{Maturo}, H.~{Al-Hraishawi}, and
  S.~{Chatzinotas}, ``{Carrier Aggregation in Satellite Communications: Impact
  and Performance Study},'' \emph{IEEE Open J. Commun. Soc.}, pp. 1--1, 2020.

\bibitem{Branch}
\BIBentryALTinterwordspacing
D.~R. Morrison, S.~H. Jacobson, J.~J. Sauppe, and E.~C. Sewell,
  ``{Branch-and-bound algorithms: A survey of recent advances in searching,
  branching, and pruning},'' \emph{Discrete Optim.}, vol.~19, pp. 79--102,
  2016. [Online]. Available:
  \url{https://www.sciencedirect.com/science/article/pii/S1572528616000062}
\BIBentrySTDinterwordspacing

\bibitem{Outer}
P.~Bonami, M.~Kilin{\c{c}}, and J.~Linderoth, ``{Algorithms and Software for
  Convex Mixed Integer Nonlinear Programs},'' in \emph{Mixed Integer Nonlinear
  Programming}, J.~Lee and S.~Leyffer, Eds.\hskip 1em plus 0.5em minus
  0.4em\relax New York, NY: Springer New York, 2012, pp. 1--39.

\bibitem{cioffi2000}
W.~Rhee and J.~M. Cioffi, ``{Increase in capacity of multiuser OFDM system
  using dynamic subchannel allocation},'' in \emph{VTC2000-Spring. 2000 IEEE
  51st Vehicular Technology Conference Proceedings (Cat. No. 00CH37026)},
  vol.~2.\hskip 1em plus 0.5em minus 0.4em\relax IEEE, 2000, pp. 1085--1089.

\bibitem{UMTS}
``{Satellite Earth Stations and Systems (SES); Satellite Component of
  UMTS/IMT2000;G-family; Part 4: Physical layer procedures},'' ETSI, European
  Standard (EN) TS 101 851-4, 2006.

\bibitem{GuidDVB}
``{Digital Video Broadcasting (DVB); Interaction channel for Satellite
  Distribution Systems; Guidelines for the Use of EN 301 790 in Mobile
  Scenarios},'' ETSI, European Standard (EN) TR 102 768, 2009.

\bibitem{itDVB}
``{Digital Video Broadcasting (DVB); Interaction channel for satellite
  distribution systems},'' ETSI, European Standard (EN) EN 301 790, 2009.

\bibitem{mosek}
\BIBentryALTinterwordspacing
M.~ApS, \emph{The MOSEK optimization toolbox for MATLAB manual. Version 9.2.},
  2021. [Online]. Available:
  \url{https://docs.mosek.com/9.2/toolbox/index.html}
\BIBentrySTDinterwordspacing

\bibitem{cvx}
M.~Grant and S.~Boyd, ``{{CVX}: Matlab Software for Disciplined Convex
  Programming, version 2.2},'' \url{http://cvxr.com/cvx}, Jan. 2020.

\bibitem{GA_Int}
\BIBentryALTinterwordspacing
K.~Deep, K.~P. Singh, M.~Kansal, and C.~Mohan, ``{A real coded genetic
  algorithm for solving integer and mixed integer optimization problems},''
  \emph{Appl. Math. Comput.}, vol. 212, no.~2, pp. 505--518, 2009. [Online].
  Available:
  \url{https://www.sciencedirect.com/science/article/pii/S0096300309001830}
\BIBentrySTDinterwordspacing

\end{thebibliography}
\bibliographystyle{IEEEtran}

\end{document}